\documentclass[%
 reprint,
 superscriptaddress,
 groupedaddress,
 showpacs,preprintnumbers,
 bibnotes,
 amsmath,amssymb,
 aps,
 prb,
 floatfix,
]{revtex4-1}

\usepackage{graphicx}
\usepackage{dcolumn}
\usepackage{bm}

\usepackage{color}

\begin{document}

\title{Crystal and electronic structure of BiTeI, AuTeI, and PdTeI compounds: A dispersion-corrected density-functional study}
\author{S\"{u}meyra G\"{u}ler-K{\i}l{\i}\c{c}}
\email{sumeyra@gtu.edu.tr} 
\affiliation{
Department of Physics, Gebze Technical University, Gebze, Kocaeli 41400, Turkey
}
\author{ \c{C}etin K{\i}l{\i}\c{c} }
\email{cetin\_kilic@gtu.edu.tr} 
\affiliation{
Department of Physics, Gebze Technical University, Gebze, Kocaeli 41400, Turkey
}

\begin{abstract}
\centerline{\sl Published version available at \url{http://dx.doi.org/10.1103/PhysRevB.91.245204}}
\vspace*{6pt}
Semilocal and dispersion-corrected density-functional calculations have been performed
  to study the crystal structure, equation of state, and electronic structure
  of metal tellurohalides with chemical formula MeTeI where Me=Bi, Au, or Pd.
A comparative investigation of the results of these calculations is conducted,
  which reveals the role of van der Waals attraction.
It is shown that the prediction of crystal structure of metal tellurohalides
  is systematically improved thanks to the inclusion of van der Waals dispersion.
It is found for BiTeI and AuTeI that
  the energy versus volume curve is anomalously flat in the vicinity of equilibrium volume and
  the calculated equation of state has an excessively steep slope in the low-pressure region,
  which are also fixed in the dispersion-corrected calculations.
Analysis based on the computation of
  the volume and axial compressibilities
  shows that predicting the anisotropy of BiTeI via the semilocal calculations
  yields an {\it unrealistic} result
  whereas the results of dispersion-corrected calculations agree with the experimental compressibility data.
Our calculations render that 
  BiTeI (AuTeI) is a narrow band gap semiconductor
  with Rashba-type spin-splitting at the band edges
  (with an indirect band gap)
  while PdTeI is a metal with relatively low density of states at the Fermi level.
The band gaps of BiTeI and AuTeI obtained via semilocal (dispersion-corrected) calculations
  are found to be {\it greater} ({\it smaller}) than the respective experimental values,
  which is against (in line with) the expected trend.
Similarly, the Rashba parameters of BiTeI are {\it bracketed} by the respective values
  obtained via semilocal and dispersion-corrected calculations,
  e.g., a larger value for the Rashba parameter $\alpha_R$ is obtained
  in association with the reduction of the band gap caused by modification of the crystal structure
  owing to van der Waals attraction.
Excellent agreement with the experimental Rashba parameters is obtained
  via interpolation of the calculated (semilocal and dispersion-corrected) values.
\end{abstract}


\maketitle


%
%
\section{\label{giris}Introduction}

\begin{figure*}
  \begin{center}
    \resizebox{1.0\textwidth}{!}{%
      \includegraphics{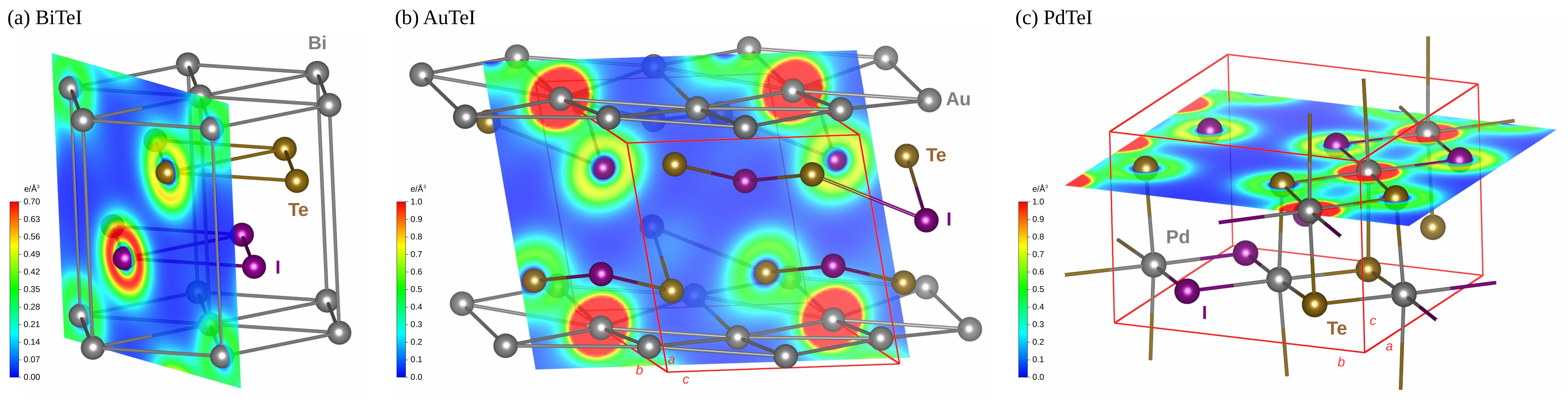}
    }
  \end{center}
  \vspace*{-0.5cm}
  \caption{(Color online)
           The crystal structures of BiTeI (a), AuTeI (b), and PdTeI (c) compounds.
           A color contour plot of the computed electronic charge density is also shown for each compound.
          }
   \label{atostr}
\end{figure*}

Bismuth tellurohalide (BiTeI)
 has recently been attracting a great deal of interest
  as a potential spintronic material
  due to emergence of giant Rashba-type spin-splittings in its band structure\cite{ishizaka11} and
                   of a pressure-induced topological insulating phase,\cite{bahramy12,xi13}
  albeit it has also been pointed out\cite{tran14} that the latter would probably be hindered by a structural phase transition.
Prompted by the discovery\cite{ishizaka11} of Rashba splittings in the band structure of BiTeI,
  density-functional calculations have increasingly been employed\cite{bahramy11,eremeev12,kulbachinskii12,sklyadneva12,fu13,rusinov13,zhu13,kanou13,wang13,chen13}
  to explore the electronic and crystal structure of this semiconductor.
Curiously, 
  although BiTeI is often described to have a layered structure
  where the triple Bi-Te-I layers are stacked along an axis perpendicular to the layers,
  the van der Waals attractions were {\it not} explicitly included in these studies.
The lattice parameters of BiTeI were therefore needed to be fixed to their experimental values\cite{shevelkov95}
  in most
  density-functional calculations.
In order to provide a more complete description,
  we adopt here the dispersion-corrected density-functional (DFT-D2) approach,\cite{grimme06}
  where the van der Waals forces are incorporated by means of a semiempirical force field.
We employ the DFT-D2 calculations for not only BiTeI but also AuTeI and PdTeI
  as far as we are aware of few experimental studies\cite{rabenau70,fenner78,seo98} on the structure and properties of the latter,
  which have not been explored via first-principles methods before.

In order to reveal the effect of the dispersion correction on the crystal and electronic structure of MeTeI (Me=Bi, Au, or Pd) compounds,
  we carried out our calculations at two levels of approximations:
  the density-functional calculations performed within the generalized gradient approximation (GGA) 
  using the functional of Perdew, Burke, and Ernzerhof (PBE)\cite{perdew96}
  without and with the semiempirical dispersion correction.
Hence, we compare the results of the PBE (semilocal) and PBE-D2 (dispersion-corrected) calculations to each other and to the experimental data.
The total energy calculations performed for compressed and dilated systems are used
  to derive equation of state for metal tellurohalides under consideration.
Our results show that inclusion of van der Waals interactions is necessary for an adequate physical description of these compounds.
In particular, predicting the anisotropy of BiTeI via the PBE calculations
  yields an {\it unrealistic} result
  whereas the results of the PBE-D2 calculations agree with the experimental compressibility data.

\begin{table}[b]
\caption{\label{D2par}
          Dispersion coefficients $C_6$ (in J nm$^6$ mol$^{-1}$) and van der Waals radii $R_0$ (in \AA)
            employed in the present PBE-D2 calculations.
          These values are taken from the references given in the rightmost column.
        }
\begin{ruledtabular}
\begin{tabular}{lccc}
Element & $C_6$ & $R_0$ & Reference             \\ \hline
Pd      & 24.67 & 1.639 & \onlinecite{grimme06} \\
Te      & 31.74 & 1.892 & \onlinecite{grimme06} \\
I       & 31.50 & 1.892 & \onlinecite{grimme06} \\
Au      & 40.62 & 1.772 & \onlinecite{amft11}   \\
Bi      & 63.55 & 1.900 & \onlinecite{clay12}   
\end{tabular}
\end{ruledtabular}
\end{table}

Although oxidation state of Bi, Au, and Pd atoms is the same ($+3$)
  in the BiTeI, AuTeI, and PdTeI compounds,
  the crystal structures of the latter are quite different
  as shown in Figs.~\ref{atostr}(a)-(c).
BiTeI, AuTeI, and PdTeI crystallize in trigonal,
                                       monoclinic, and 
                                       tetragonal structures
  with space groups $P3m1$ (No. 156), 
                    $P2_1/c$ (No. 14), and
                    $P4_2/mmc$ (No. 131), respectively.
The crystal structure of BiTeI could be characterized by the hexagonal lattice parameters
  $a$ and $c$, and two internal parameters
  since Bi, Te, and I atoms occupy the $1a$, $1c$, and $1b$ positions
  with fractional coordinates (0,0,0),
                              (2/3,1/3,$z_{\rm Te}$), and
                              (1/3,2/3,$z_{\rm I}$), respectively,
  cf. Ref.~\onlinecite{shevelkov95}.
As shown in Figure~\ref{atostr}(a),
  adjacent trigonal layers formed by Bi, Te, and I atoms stack along the $c$-axis
  of the hexagonal lattice in BiTeI.
The crystal structure of AuTeI could be characterized by the monoclinic lattice parameters $a$, $b$, $c$, and $\beta$,
  and nine internal parameters
  since Au, Te, and I atoms occupy the $4e$ positions
  with fractional coordinates ($x_{\rm Au}$,$y_{\rm Au}$,$z_{\rm Au}$),
             ($x_{\rm Te}$,$y_{\rm Te}$,$z_{\rm Te}$),
             ($x_{\rm I }$,$y_{\rm I }$,$z_{\rm I }$), respectively,
  cf. Ref.~\onlinecite{fenner78}.
As shown in Fig.~\ref{atostr}(b),
  the atoms form corrugated layers parallel to the $bc$-plane,
  which stack along the $a$-axis in AuTeI.
The crystal structure of PdTeI could be characterized by the tetragonal lattice parameters $a$ and $c$,
  and three internal parameters
  since Pd, Te, and I atoms occupy the $4m$, $4l$, and $4j$ positions
  with fractional coordinates (0,$y_{\rm Pd}$,0),
                              ($x_{\rm Te}$,0,0), and 
                              ($x_{\rm I }$,1/2,0), respectively,
  cf. Ref.~\onlinecite{seo98}.
As shown in Fig.~\ref{atostr}(c),
  ladder chains made of  Te$_2$PdI$_2$ units running along the $a$- and $b$-axes
  stack along the $c$-axis in PdTeI.
Note that the Te and I atoms are mixed in the layers or ladder chains of AuTeI or PdTeI, respectively,
  whereas the layers of BiTeI are unary, consisting of Bi, Te or I atoms.
Accordingly, the space group of the BiTeI crystal is noncentrosymmetric
  while the crystal structures of AuTeI and PdTeI are centrosymmetric.
This puts BiTeI in a unique position in terms of spintronic material properties,\cite{ishizaka11}
  which lacks inversion symmetry in the crystal structure.

The rest of the paper is organized as follows:
Sec.~\ref{yontem} is devoted to the method of calculation, giving also a summary of the computational details.
We focus on the crystal structure, equation of state, and electronic
  band  
  structure in Sec.~\ref{sonuclar}
  where we discuss the results of our semilocal (PBE) or dispersion-corrected (PBE-D2) density-functional calculations.
A summary of our findings is given in Sec.~\ref{netice}.
 Lastly, the spin-orbit coupling (SOC) effects on the BiTeI band structure are described in further details in the Appendix~\ref{SOetkisi}. 
%
%
\section{\label{yontem}Method}

All calculated properties reported here were obtained
  via semilocal\cite{perdew96} (PBE) or dispersion-corrected\cite{grimme06} (PBE-D2) density-functional calculations
  performed by employing the projector augmented-wave (PAW) method,\cite{blochl94} as implemented\cite{kresse99}
  in the Vienna {\it ab initio} simulation package\cite{kresse96} (VASP).
The calculations for BiTeI and AuTeI
  were performed in the noncollinear mode\cite{hobbs00,marsman02} of VASP
  in order to take spin-orbit coupling into account.
The 4$d$ and 5$s$,
    5$s$ and 5$p$,
    5$s$ and 5$p$,
    5$d$ and 6$s$, and
    6$s$ and 6$p$
    states are treated as valence states for
    palladium,
    tellurium,
    iodine,
    gold, and 
    bismuth, respectively.
Plane wave basis sets were used to represent the electronic states, which were determined by
imposing a kinetic energy cutoff of 325 eV.
The long-range dispersion corrections\cite{grimme06} for periodic systems were treated as described in Ref.~\onlinecite{bucko10}.
The values of the dispersion coefficient $C_6$ and van der Waals radius $R_0$
  used in this study are given in Table~\ref{D2par}.
The global scaling factor $s_6$ was set to 0.75, which is the adequate value for the PBE functional.\cite{bucko10}

We first carried out full optimization of the crystal structures
  where concurrent relaxations of the unit cell volume and shape as well as the ionic positions were performed
  with no symmetry constraints,
  until the maximum value of residual forces on atoms was reduced to be smaller than 0.01 eV/\AA.
Convergence criterion for the electronic self-consistency was set up to 10$^{-6}$~eV.
In these optimizations,
  we used the primitive unit cells of BiTeI, AuTeI, and PdTeI,
  whose Brillouin zones were sampled by
  $20\times 20\times 16$,
  $8\times 8\times 8$, and
  $21\times 21\times 29$
  {\bf k}-point meshes, respectively, which were generated according to Monkhorst-Pack scheme.\cite{monkhorst76}
Using the optimized crystal structures,
  we then carried out band-structure and (projected) density-of-states calculations.
Besides, we performed geometry optimizations
  for the elemental solids of bismuth, gold, palladium, tellurium, and iodine,
  and employed the respective equilibrium energies per atom
  $E_{\rm Me}$ (Me=Bi, Au, Pd), $E_{\rm Te}$ and $E_{\rm I}$
  in the computation of the formation energy $\Delta H_f$.
It should be reminded that
  the form of the electronic Hamiltonian used in dispersion-corrected (PBE-D2) calculations
  is the same as in the calculations employing the PBE functional alone.
In other words,
  the effect of dispersion correction on the electronic structure
  is {\it indirectly} through modification of the crystal structure
  since the van der Waals interactions are treated as semiempirical force fields
  in the DFT-D2 approach.\cite{grimme06}

Secondly, we carried out constant-volume optimization of the crystal structures
  where the unit cell shape and the ionic positions were allowed to relax.
Hence, we obtained the energy $E$ per formula unit
  as a function of the volume $V$ per formula unit,
  which was used to derive equation of state (EOS) at zero temperature.
We found that 
  the energy-volume curve is not accurately reproduced
  by a third-order Birch-Murnaghan (BM) fit that is in widespread use,
  which is further discussed in Section~\ref{sonuclar}.
Thus, we performed forth- and fifth-order BM fits\cite{angel00,strachan99}
  as a function of the Eulerian strain $f= [ ( V_0/V  )^{2/3} - 1 ]/2$
  that is defined from $V$ and the zero-pressure volume $V_0$, employing
  \begin{equation}
    E=E_0+\sum_{k=2}^{k_{\rm max}} C_{k-1} f^{k} 
    \label{BM45ene}
  \end{equation}
  with $k_{\rm max}=$ 4 and 5, respectively.
Here $C_k$s are the fitting coefficients, and
  $E_0$ denotes the equilibrium energy (per formula unit).
Note that $\Delta H_f=E_0-(E_{\rm Me}+E_{\rm Te}+E_{\rm I})$.
The pressure $P$ was computed by using
  \begin{equation}
    P=\frac{(1+2f)^{\frac{5}{2}}}{3V_0} \sum_{k=1}^{k_{\rm max}-1} (k+1) C_k f^k. 
    \label{BM45pre}
  \end{equation}
The isothermal bulk modulus $K_0$ and its pressure derivatives
  $K^\prime_0$, $K^{\prime\prime}_0$, and $K^{\prime\prime\prime}_0$
  (all evaluated at $V_0$)
  were obtained via
  \begin{eqnarray}
    K_0                     &=&\frac{2C_1}{9V_0}, \nonumber \\ 
    K^\prime_0              &=&\frac{C_2}{C_1}+4, \nonumber \\ 
    K^{\prime\prime}_0      &=&\frac{1}{K_0}\left ( \frac{4C_3}{3C_1} - K^\prime_0(K^\prime_0-7) -\frac{143}{9} \right ), \nonumber \\ 
    K^{\prime\prime\prime}_0&=&\frac{1}{9K_0^2} \left ( \frac{20C_4}{C_1} -12K_0(3K^\prime_0-8)K^{\prime\prime}_0 \right. \nonumber \\ 
                            & & \left. - K^\prime_0[(3K^\prime_0-16)^2+118]+\frac{1888}{3} \right ), 
    \label{BM45bm}
  \end{eqnarray}
respectively.

The volume compressibility $\kappa_v$ was obtained as the inverse of the bulk modulus, i.e., $\kappa_v=1/K_0$.
The axial (linear) compressibilities $\kappa_l=-\left ( d \ln l / dP \right )_{P=0}$,
  with $l$ denoting the lattice constant along one of the crystal axes,
  were computed
  by dividing $-\frac{1}{l}\left ( \frac{dl}{dV} \right )_{V=V_0}$ by $\left ( \frac{dP}{dV}\right )_{V=V_0}$,
  where the former [latter] was
  obtained via cubic spline interpolation of the lattice parameter $l$ as a function of the volume $V$
              [via Eq.~(\ref{BM45pre})].
The reliability of this procedure was tested
  by checking whether $\kappa_v=\kappa_a+\kappa_b+\kappa_c$ holds or not.
The latter equality was satisfied
  in the cases of BiTeI and PdTeI,
  but not in the case of AuTeI.
As discussed in Section~\ref{sonuclar},
  the variation of the lattice parameter $c$ with the pressure is nonmonotonic in the case of AuTeI,
  which results in a substantial error in estimating derivatives
  via spline interpolation.
The linear compressibilities of AuTeI are therefore not reported.

%
%
\section{\label{sonuclar}Results and Discussion}

%
%
\subsection{Crystal Structure}

The crystal structure optimizations
  result in the experimentally determined ground-state structures,
  i.e., a noncentrosymmetric (centrosymmetric) crystal for BiTeI (AuTeI and PdTeI)
  that are shown in Fig.~\ref{atostr}(a)-(c)
  where color contour plots of computed electronic charge density are also displayed.
Inspection of the charge-density plot in Fig.~\ref{atostr}(a)
 reveals that the most electron-rich (electron-poor) regions in BiTeI are around I (Bi) atoms.
This reflects the fact that bismuth is less electronegative than both iodine and tellurium.\cite{CRC}
It is also interesting to note that
  the electronegativity difference $\chi_{\rm Te}-\chi_{\rm Bi}$ is {\it positive} and comparatively {\it small}
  whereas $\chi_{\rm I}-\chi_{\rm Bi}$ is also {\it positive} but comparatively {\it large}.\cite{eleneg,CRC}
Accordingly,
  the electronic charge distribution around Bi$-$Te (Bi$-$I) bonds in Fig.~\ref{atostr}(a)
  is rather of a covalent (ionic) character,
  which introduces an asymmetry between Bi$-$Te and Bi$-$I bonds.
On the other hand, we have
$\chi_{\rm Te} - \chi_{\rm Me} < 0 < \chi_{\rm I}  - \chi_{\rm Me} < \chi_{\rm I}  - \chi_{\rm Te}$
  for both Me=Au and Pd,
  i.e., the differences $\chi_{\rm Te} - \chi_{\rm Me}$ and $\chi_{\rm I}  - \chi_{\rm Me}$ are both relatively {\it small},
  which are of the {\it opposite} sign.\cite{eleneg,CRC}
In agreement with the latter,
  the electronic charge distribution around {\it not only} Me$-$Te bonds and {\it but also} Me$-$I bonds in Fig.~\ref{atostr}(b)-(c)
  look more like those of covalent bonds.
Thus, the tellurium and iodine atoms
  prefer to coordinate with the metal atoms {\it almost equally},
  and therefore tend to mix (as opposed to form unary layers) in the AuTeI and PdTeI compounds,
  rendering the crystal structures of the latter centrosymmetric.

\begin{figure}
  \begin{center}
    \resizebox{0.5\textwidth}{!}{%
      \includegraphics{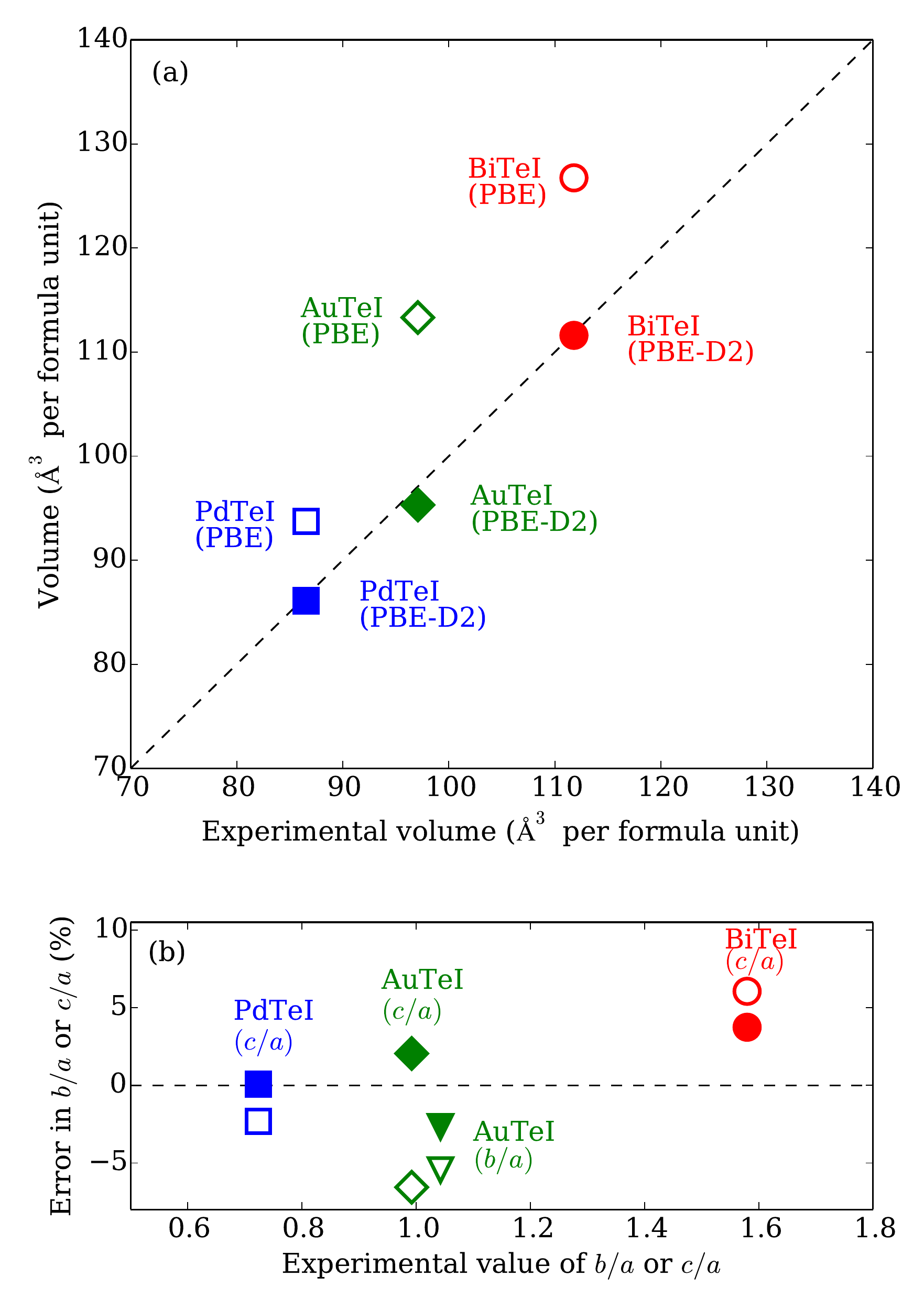}
    }
  \end{center}
  \vspace*{-0.5cm}
  \caption{(Color online) The plots of
           calculated versus experimental equilibrium volume (a), and
           error in b/a or c/a ratios (b).
          }
   \label{hacim}
\end{figure}

A comparison of the results of the crystal structure optimizations (PBE and PBE-D2)
   to the experimental data is given in Tables~\ref{tabcs} and S1 (Ref.~\onlinecite{supmat}).
It is seen in Table~\ref{tabcs} that the improvement due to the dispersion correction
  is mostly on the lattice parameters ($a$, $b$, $c$)
  whereas the PBE and PBE-D2 calculations yield errors of similar magnitude
  in the prediction of the internal parameters ($x$, $y$, $z$).
It is thus notable that the prediction of unit cell volume and shape
  is substantially improved:
Figure~\ref{hacim}(a) displays
  a plot of the calculated (PBE and PBE-D2) versus experimental values for the equilibrium volume per formula unit.
Although it is well known the unit cell volume is {\it overestimated} within the GGA,
  the error in the PBE-optimized volume is clearly greater than expected,
  which is 13.4, 16.7, and 8.3 \% for BiTeI, AuTeI, and PdTeI, respectively.
As evident from the trend of the empty symbols in Fig.~\ref{hacim}(a)
  this anomalous overestimation is widespread.
Since the lattice parameters ($a$, $b$, $c$) obtained in the PBE-D2 optimizations
  are {\it smaller} than those in the PBE-D2 optimizations, cf. Table~\ref{tabcs},
  the prediction of equilibrium volume is systematically improved thanks to the dispersion correction,
  as evident from the trend of the filled symbols in Fig.~\ref{hacim}(a).
Moreover, a comparison of the empty (PBE) and filled (PBE-D2) symbols of the same shape to each other
  in Fig.~\ref{hacim}(b)
  shows that the errors in the prediction of $b/a$ and $c/a$ ratios
  are, at the same time, significantly reduced
  thanks to the dispersion correction.

\begin{table}
\caption{\label{tabcs} Calculated (PBE and PBE-D2) and measured lattice parameters of the MeTeI crystals.
                       The experimental values are taken from Ref.~\onlinecite{shevelkov95}, Ref.~\onlinecite{fenner78}, and Ref.~\onlinecite{seo98} for BiTeI, AuTeI, and PdTeI, respectively.
        }
\begin{ruledtabular}
\begin{tabular}{llccc}
      &              &   PBE   & PBE-D2  & Exptl.  \\ \hline
BiTeI &              &         &         &         \\
      & $a$ (\AA)    & 4.4371  & 4.2843  & 4.3392  \\
      & $c$ (\AA)    & 7.433   & 7.021   & 6.854   \\
      & $z_{\rm Te}$ & 0.7692  & 0.7479  & 0.6928  \\
      & $z_{\rm I}$  & 0.2828  & 0.3115  & 0.2510  \\
AuTeI &              &         &         &         \\
      & $a$ (\AA)    & 8.0057  & 7.2579  & 7.3130  \\
      & $b$ (\AA)    & 7.8918  & 7.3654  & 7.6242  \\
      & $c$ (\AA)    & 7.4208  & 7.3483  & 7.2550  \\
      & $\beta$      & 104.81  & 103.95  & 106.26  \\
      & $x_{\rm Au}$ & 0.4589  & 0.4746  & 0.4654  \\
      & $y_{\rm Au}$ & 0.1333  & 0.1370  & 0.1395  \\
      & $z_{\rm Au}$ & 0.2308  & 0.2405  & 0.2370  \\
      & $x_{\rm Te}$ & 0.6481  & 0.6868  & 0.6720  \\
      & $y_{\rm Te}$ & 0.1433  & 0.1331  & 0.1301  \\
      & $z_{\rm Te}$ & 0.9821  & 0.9908  & 0.9910  \\
      & $x_{\rm I}$  & 0.1954  & 0.1626  & 0.1758  \\
      & $y_{\rm I}$  & 0.0659  & 0.1059  & 0.0857  \\
      & $z_{\rm I}$  & 0.3895  & 0.3799  & 0.3892  \\
PdTeI &              &         &         &         \\
      & $a$ (\AA)    & 8.095   & 7.806   & 7.821   \\
      & $c$ (\AA)    & 5.722   & 5.652   & 5.659   \\
      & $y_{\rm Pd}$ & 0.2500  & 0.2495  & 0.2525  \\
      & $x_{\rm Te}$ & 0.2086  & 0.2262  & 0.2164  \\
      & $x_{\rm I} $ & 0.2321  & 0.2462  & 0.2435
\end{tabular}
\end{ruledtabular}
\end{table}

As for the prediction of the internal parameters of the BiTeI crystal,
  it is interesting to 
  compare our results to the results of 
  Bahramy {\it et al.}\cite{bahramy11} (Sklyadneva {\it et al.}\cite{sklyadneva12})
  who predicted $z_{\rm Te}=0.7482$ and $z_{\rm I}=0.3076$ ($z_{\rm Te}=0.758$ and $z_{\rm I}=0.299$).
Note that
  our PBE-D2 and PBE values for $z_{\rm Te}$ and $z_{\rm I}$
  are in close agreement with those given
  by Bahramy {\it et al.}\cite{bahramy11} and Sklyadneva {\it et al.},\cite{sklyadneva12} respectively
  -albeit in both Refs.~\onlinecite{bahramy11} and \onlinecite{sklyadneva12}
  the PBE functional
  was employed with {\it no} additional terms for the dispersion correction.
This is so because
  the lattice parameters $a$ and $c$ were fixed to their experimental values (that are close to our PBE-D2 values)
  in Ref.~\onlinecite{bahramy11},
  which was not done in Ref.~\onlinecite{sklyadneva12}.

The predictions of Ref.~\onlinecite{bahramy11} and \onlinecite{sklyadneva12}
  show that
  the computed values of the equilibrium Bi$-$Te and Bi$-$I distances correspond to
  the Bi$-$I and Bi$-$Te distances in the experimentally determined structure,\cite{shevelkov95} respectively,
  which is supported by our PBE-D2 values, cf. Table~S1 (Ref.~\onlinecite{supmat}).
Note that in Ref.~\onlinecite{shevelkov95} the bond lengths given in Table 3 are {\it not} consistent
   with the internal parameters given Table~2.
It is therefore problematic to perform a {\it direct} comparison of
  experimental and optimized values of the internal lattice parameters $z_{\rm Te}$ and $z_{\rm I}$.
Thereupon, we considered a second ``phase'' for BiTeI,
  which was generated by {\it exchanging} the positions of Te and I atoms
  so that the Te and I atoms occupy the $1b$ and $1c$ positions
          with fractional coordinates (1/3,2/3,$z_{\rm Te}$), and
                                      (2/3,1/3,$z_{\rm I}$), respectively.
The crystal structure optimization performed for this second phase yielded
  $z_{\rm Te}=0.2520$ and $z_{\rm I}=0.6884$,
  and the total energy {\it the same as}
  that of the BiTeI phase described in Sec.~\ref{giris} and Table~\ref{tabcs}.
In line with the latter,
  comparison performed with the aid of the COMPSTRU program\cite{tasci12}
  proved that
  this second BiTeI phase (Te atoms residing at $1b$ positions with $z_{\rm Te}=0.2520$ and I atoms residing at $1b$ positions with $z_{\rm I}=0.6884$)
  is indeed {\it identical}
  to the first one        (Te atoms residing at $1c$ positions with $z_{\rm Te}=0.7479$ and I atoms residing at $1b$ positions with $z_{\rm I}=0.3115$).
It is interesting to point out that
  excellent agreement between experimental data and our computed values ($z_{\rm Te}=0.2520$ and $z_{\rm I}=0.6884$)
  is obtained {\it once} the positions of Te and I atoms are exchanged in the experimental\cite{shevelkov95} crystal structure of BiTeI.
As long as the x-ray diffraction fails to distinguish the Te and I layers in BiTeI,\cite{bahramy11} 
  we anticipate that a full-fledged experimental characterization                                  
  would render the values of $z_{\rm Te}$ and $z_{\rm I}$                                          
  in agreement with our optimized (PBE-D2) values                                                  
  given in Table~\ref{tabcs}.                                                                      

%
%
\subsection{\label{subsec:eos}Equation of State}

\begin{figure*}
  \begin{center}
    \resizebox{1.0\textwidth}{!}{%
      \includegraphics{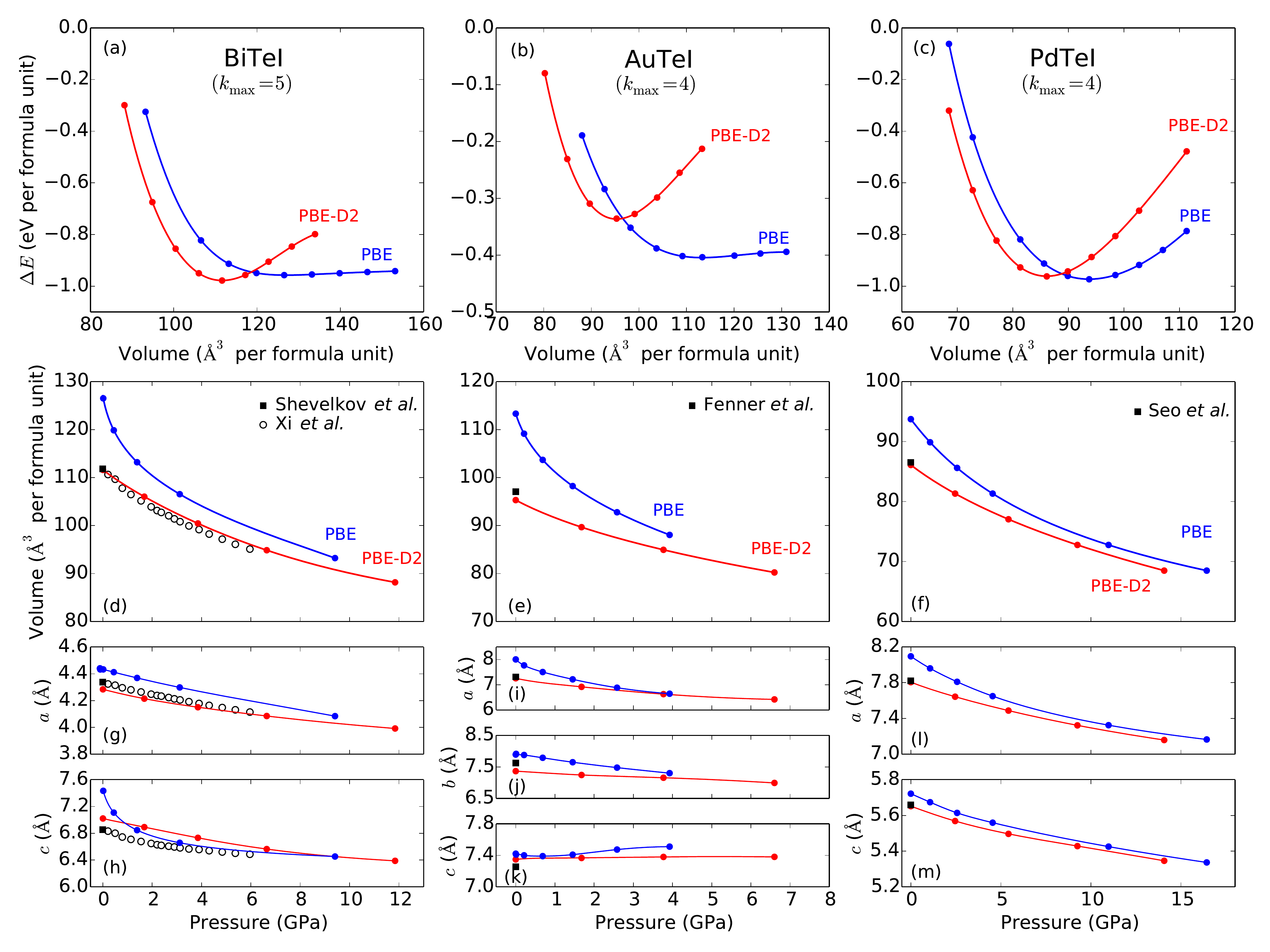}
    }
  \end{center}
  \vspace*{-0.5cm}
  \caption{(Color online)
           The plot of the energy difference $\Delta E=E-(E_{\rm Me}+E_{\rm Te}+E_{\rm I})$, cf. Eq.~(\ref{BM45ene}),
            versus the volume for BiTeI (a), AuTeI (b), and PdTeI (c).
           Here $k_{\rm max}$ is set to 4 for BiTeI, and 3 for AuTeI and PdTeI.
           The respective pressure-volume curves are plotted in (d), (e), and (f).
           The variation of the lattice parameters ($a$, $b$, $c$) with the pressure is shown in (g)-(m).
           The PBE-calculated and dispersion-corrected (PBE-D2) curves are in blue and red, respectively.
           The filled symbols mark the experimental values of the equilibrium volumes [(d)-(f)] and lattice parameters [(g)-(m)]
             measured by Shevelkov {\it et al.} (Ref.~\onlinecite{shevelkov95}),
                           Fenner {\it et al.} (Ref.~\onlinecite{fenner78}), and
                              Seo {\it et al.} (Ref.~\onlinecite{seo98})
             for BiTeI, AuTeI, and PdTeI, respectively.
           The empty circles in (d), (g), and (h) represent the experimental compressibility data
             of Xi {\it et al.} (Ref.~\onlinecite{xi13}) for BiTeI.
          }
   \label{uyarla}
\end{figure*}

\begin{table*}
\caption{\label{tabfit}
        The values of the formation energy $\Delta H_f$,
                      the equilibrium volume $V_0$, and 
                      the fitting coefficients $C_k$ in Eq.~(\ref{BM45ene})
                      for MeTeI (Me=Bi,Au,Pd) compounds.
        }
\begin{ruledtabular}
\begin{tabular}{llccrrrrrr}
      &              & $k_{\rm max}$ & $\Delta H_f$ (eV) & $V_0$(\AA$^3$) & $C_1$ (eV) & $C_2$ (eV) & $C_3$ (eV) & $C_4$ (eV)  \\ \hline
BiTeI &              &               &                         &                &            &            &            &             \\
      & PBE          & 5             & -0.958                  & 127.039        &  15.357    & 278.613    &  1091.512  & -8932.044   \\
      & PBE-D2       & 5             & -0.978                  & 111.626        &  88.806    & 228.317    & -4645.020  & 30756.053   \\
AuTeI &              &               &                         &                &            &            &            &             \\
      & PBE          & 4             & -0.404                  & 113.317        &  13.558    & 172.994    &  -444.287  &             \\  
      & PBE-D2       & 4             & -0.337                  &  95.308        &  58.526    & 244.128    & -1103.027  &             \\  
PdTeI &              &               &                         &                &            &            &            &             \\
      & PBE          & 4             & -0.973                  &  93.742        &  61.728    & -11.560    &   506.408  &             \\  
      & PBE-D2       & 4             & -0.961                  &  86.101        &  90.574    & 103.902    &  -695.661  &                 
\end{tabular}
\end{ruledtabular}
\end{table*}

\begin{figure}
  \begin{center}
    \resizebox{0.5\textwidth}{!}{%
      \includegraphics{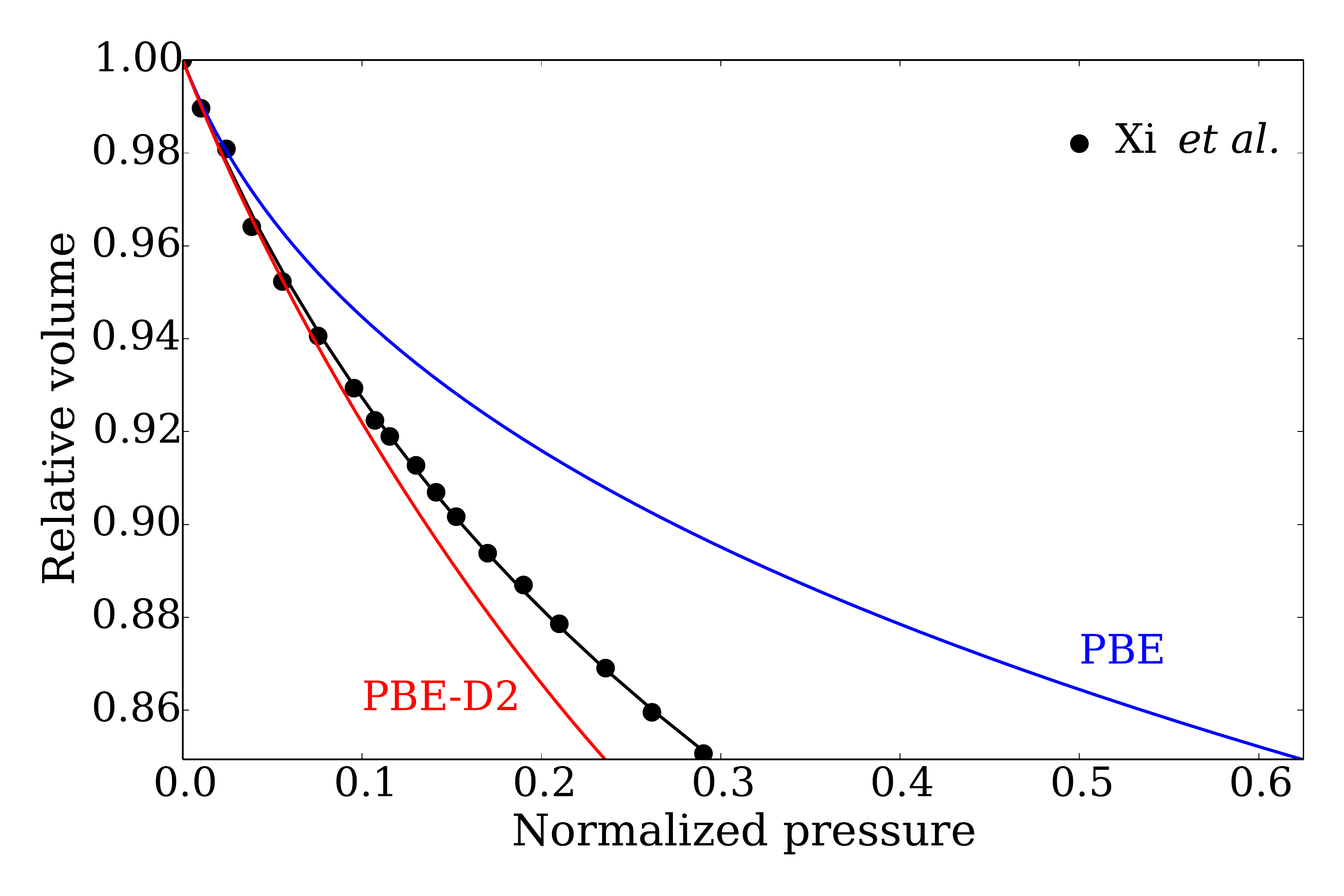}
    }
  \end{center}
  \vspace*{-0.7cm}
  \caption{(Color online)
           The plot of the relative volume $V/V_0$ versus the normalized pressure $P/K_0$ for BiTeI.
           The PBE-calculated and dispersion-corrected curves are in blue and red, respectively.
           The black (solid) circles represent the experimental data of Xi {\it et al.} (Ref.~\onlinecite{xi13}).
           The black curve was obtained by performing a third-order BM fit to the latter.
          }
   \label{kunc}
\end{figure}

\begin{table}[b]
\caption{\label{tabbm}
        The bulk modulus $K_0$ (in GPa) and its pressure derivatives $K_0^{\prime}$, $K_0^{\prime\prime}$ (in GPa$^{-1}$), and $K_0^{\prime\prime\prime}$ (in GPa$^{-2}$).
        }
\begin{ruledtabular}
\begin{tabular}{llcrrrr}
      &              & $k_{\rm max}$  & $K_0$    & $K_0^{\prime}$ & $K_0^{\prime\prime}$ & $K_0^{\prime\prime\prime}$ \\ \hline
BiTeI &              &                &          &                &                      &                            \\
      & PBE          & 5              &  4.3     & 22.1           & -59.6                & 658.8                      \\
      & PBE-D2       & 5              & 28.3     &  6.6           &  -2.9                &   2.5                      \\
      & Experimental & 3              & 20.5     &  7.6           &                      &                            \\
AuTeI &              &                &          &                &                      &                            \\
      & PBE          & 4              &  4.3     & 16.8           & -52.4                &                            \\
      & PBE-D2       & 4              & 21.9     &  8.2           &  -2.3                &                            \\
PdTeI &              &                &          &                &                      &                            \\
      & PBE          & 4              & 23.4     &  3.8           &  0.3                 &                            \\
      & PBE-D2       & 4              & 37.5     &  5.1           & -0.4                 &                            
\end{tabular}
\end{ruledtabular}
\end{table}

\begin{table}
\caption{\label{tabac}
        The volume ($\kappa_v$) and axial ($\kappa_a$ and $\kappa_c$) compressibilities (in GPa$^{-1}$),
        and their ratios.
        }
\begin{ruledtabular}
\begin{tabular}{llcccccc}
      &              & $\kappa_v$ & $\kappa_a$ & $\kappa_c$ & $\kappa_a/\kappa_v$ & $\kappa_c/\kappa_v$ & $\kappa_a/\kappa_c$ \\ \hline
BiTeI &              &            &            &            &                     &                     &                     \\
      & PBE          & 0.232      & 0.010      & 0.212      & 0.043               & 0.914               & 0.047               \\
      & PBE-D2       & 0.035      & 0.011      & 0.013      & 0.314               & 0.371               & 0.846               \\
PdTeI &              &            &            &            &                     &                     &                     \\
      & PBE          & 0.043      & 0.017      & 0.008      & 0.395               & 0.186               & 2.125               \\
      & PBE-D2       & 0.027      & 0.010      & 0.007      & 0.370               & 0.259               & 1.429               
\end{tabular}
\end{ruledtabular}
\end{table}

A plot of the energy difference $\Delta E=E-(E_{\rm Me}+E_{\rm Te}+E_{\rm I})$
  with respect to the volume $V$
  is given in Figs.~\ref{uyarla}(a), (b), and (c) for BiTeI, AuTeI, and PdTeI, respectively.
The calculated values are represented by the blue (PBE) and red (PBE-D2) circles.
The solid-line curves connecting the symbols show the forth- and fifth-order BM fits,
  whose equation is given by $\Delta E=\Delta H_f+\sum_{k=2}^{k_{\rm max}} C_{k-1}f^k/k$,
  cf. Eq.~(\ref{BM45ene}).
Although we first performed the BM fits with $k_{\rm max}=3$,
  we found it necessary to increase $k_{\rm max}$ to 4 for AuTeI and PdTeI and to 5 for BiTeI
  in order to ensure a satisfactory level of accuracy in the fitting procedure,
  as mentioned in Section~\ref{yontem}.
It is discernible in Fig.~S1 (Ref.~\onlinecite{supmat}) that
  a good fit is not obtained when $k_{\rm max}$ is reduced to 4 for BiTeI, and 3 for AuTeI and PdTeI.

The values of $C_k$ parameters obtained via fitting
  as well as the formation energies $\Delta H_f$ are given in Table~\ref{tabfit}.
Note that
  while $\Delta H_f$ is significantly (AuTeI) or slightly (PdTeI) reduced
  due to addition of the dispersion terms to the PBE functional,
  both PBE and dispersion-corrected (PBE-D2) calculations
  yield a similar degree of overbinding for BiTeI
  since the experimental value\cite{aliev08} of the enthalpy of formation for BiTeI
  is $\Delta H_f^0 (298~{\rm K})= 0.856$ eV per formula unit.
On the other hand,
  the interlayer binding energy of BiTeI
  was computed to be 24 and 332 meV per formula unit in our PBE and PBE-D2 calculations,
  respectively.
%
%
%
%
%
Thus, the dispersion-corrected (PBE-D2) calculations
  yield a substantially stronger interlayer binding for BiTeI,
  compared to the PBE calculations.
It is also notable that the PBE-calculated binding curves of BiTeI [Fig.~\ref{uyarla}(a)] and AuTeI [Fig.~\ref{uyarla}(b)]
  are anomalously flat in the vicinity of equilibrium volume $V_0$
  (and especially for volumes larger than $V_0$),
  which is not the case for PdTeI [Fig.~\ref{uyarla}(c)].
This flatness of the binding curves
  could be attributed to lacking van der Waals interactions
  because
  the dispersion-corrected (PBE-D2) binding curves have a significantly increased curvature.

The pressure-volume curves are shown
  in Figs.~\ref{uyarla}(d), (e), and (f) 
  for BiTeI, AuTeI, and PdTeI, respectively.
The graphs given in Figs.~\ref{uyarla}(g)-(m)
  shows the variation of the lattice parameters ($a$, $b$, $c$) with the pressure.
The experimental compressibility data\cite{xi13}
  is also included in Figs.~\ref{uyarla}(d), (g), and (h)
  for comparison.
The experimental data employed in Fig.~5 of Ref.~\onlinecite{chen13}
  is not presented in Fig.~\ref{uyarla}(d)
  since is seems to be shifted in comparison to the data of Ref.~\onlinecite{xi13},
  cf. Fig.~S2 (Ref.~\onlinecite{supmat}),
  which also does not agree with the equilibrium volume reported in Ref.~\onlinecite{shevelkov95}.
It is noticeable in Fig.~\ref{uyarla}(d) that
  the dispersion-corrected $V(P)$ curve is
  in much better agreement with the experimental data (the empty circles)
  whereas the PBE-calculated curve has a substantially larger slope
  in the low-pressure region.
The latter appears to be the case with AuTeI too, cf. Fig.~\ref{uyarla}(e).
In contrast, the PBE-calculated and dispersion-corrected (PBE-D2) curves for PdTeI
  have comparable slopes even in the low-pressure region.
Note that the steeper decay of $V(P)$ curve in the low-pressure region
  could be traced back to the variation of lattice parameters $c$ and $a$ for
  BiTeI and AuTeI, respectively, cf. Figs.~\ref{uyarla}(h) and (i).
Thereupon, the error in the PBE calculations
  is clearly not restricted to the overestimation of the volume
  for quasi-layered systems, e.g., Figs.~\ref{atostr}(a) and (b).
One should, on the other hand, also notice that
  the agreement between the PBE-D2 and experimental curves in Fig.~\ref{uyarla}(d) is rather coincidental
  since the lattice parameter $c$ ($a$) is overestimated (underestimated) in the PBE-D2 calculations,
  regardless the value of $P$,
  cf. Figs.~\ref{uyarla}(g) and (h).
Furthermore, the calculated and experimental values in Fig.~\ref{uyarla}
  refer to zero and room temperature, respectively.
Therefore, a comparison of the relative volume ($V/V_0$) versus normalized pressure ($P/K_0$) curves
  is given in Fig.~\ref{kunc} for BiTeI,
  following a proposal of Ref.~\onlinecite{kunc10}
  [where using a simple {\it scaling} of variables was demonstrated to be practical
  in comparing the calculated $P(V)$ equation to the experimental equation of state].
It is seen in Fig.~\ref{kunc} that the experimental data points
  are {\it bracketed} by the PBE-calculated and dispersion-corrected (PBE-D2) curves,
  which are seemingly in better agreement with the PBE-D2 curve than the PBE curve.

Turning back to Fig.~\ref{uyarla},
  it is interesting to point out that
  the variation of the lattice parameter $c$ of AuTeI with the pressure
  is slightly increasing (nonmonotonic) according to the results of the PBE-D2 (PBE) calculations,
  as shown in Fig.~\ref{uyarla}(k).
This implies that the lattice of AuTeI would expand along the $c$-axis under compression.
In other words,
  the linear compressibility $\kappa_c$ of AuTeI is {\it negative}
  according to our calculations, cf. Fig.~\ref{uyarla}(k).                                     
It should be remarked that this prediction calls for experimental verification                 
  inasmuch as the compressibility of AuTeI has not been investigated before, to our knowledge. 
Note that the usual behavior under compression is that the individual lattice parameters
  decrease so that $\kappa_l>0$,
  which is the case in Figs.~\ref{uyarla}(g)-(j) and (l)-(m).
The 
  behavior of {\it negative linear compressibility} has
  nevertheless
  been observed in a number of systems\cite{goodwin08,fortes11,li12,cairns13} recently.

The bulk modulus $K_0$ and
  its pressure derivatives $K_0^{\prime}$, $K_0^{\prime\prime}$, and $K_0^{\prime\prime\prime}$
  computed via Eq.~(\ref{BM45bm}) with $C_k$ coefficients listed in Table~\ref{tabfit}
  are given in Table~\ref{tabbm}
  where the experimental values for BiTeI are also included.
The latter were obtained by performing a third-order BM fit, cf. the black solid curve in Fig.~\ref{kunc},
  to the experimental compressibility data.\cite{xi13}
It is seen that
  the PBE calculation yields a substantially underestimated (overestimated) value
  for $K_0$ ($K_0^\prime$) of BiTeI.
Recalling that
  (i) the PBE-calculated binding energy curve anomalously flat in the vicinity of equilibrium volume, cf. Fig.~\ref{uyarla}(a), and 
  (ii) the PBE-calculated equation of state has an excessively steep slope in the low-pressure region, cf. Fig.~\ref{uyarla}(d),
  it is no surprise that the bulk modulus  is substantially underestimated in the PBE calculations.
The PBE-calculated first pressure derivative $K_0^\prime$ is consequently greatly overestimated,
  balancing this underestimation of $K_0$.
On the contrary, the dispersion-corrected (PBE-D2) calculations,
  albeit in much better agreement with the experimental values,
  result in overestimation (underestimation) for $K_0$ ($K_0^\prime$).
Hence,
  the experimental values of $K_0$ and $K_0^\prime$
  are {\it bracketed} by the PBE-calculated and dispersion-corrected (PBE-D2) values.

The volume (bulk) and axial (linear) compressibilities
  and their ratios are given in Table~\ref{tabac} for BiTeI and PdTeI.
Note that $\kappa_b=\kappa_a$ for these systems.
For BiTeI (PdTeI), the PBE and PBE-D2 values for the $a$-axis ($c$-axis) compressibility
  are close to each other.
In contrast, $\kappa_c^{\rm PBE} \ll \kappa_c^{\rm PBE-D2}$ for BiTeI,
         and $\kappa_a^{\rm PBE}   > \kappa_a^{\rm PBE-D2}$ for PdTeI.
Thus, the $c$-axis compressibility of BiTeI is greatly overestimated in the PBE calculations
  (which reflects lacking van der Waals binding).
We have furthermore
  $\kappa_a/\kappa_c \ll 1 $ (PBE) while 
  $\kappa_a/\kappa_c \sim 1$ (PBE-D2), and
  $\kappa_a/\kappa_v < 1/3 $ and  $\kappa_c/\kappa_v > 1/3 $ (PBE) while
  $\kappa_a/\kappa_v \sim 1/3 $ and  $\kappa_c/\kappa_v \sim 1/3 $ (PBE-D2)
  for BiTeI.
Note that the ratio $\kappa_a/\kappa_c$ should be unity, and 
  both $\kappa_a/\kappa_v$ and $\kappa_c/\kappa_v$ need to be equal to 1/3
  for {\it isotropic} materials.
For {\it layered} materials, on the other hand, 
  one would expect to have 
  the ratio $\kappa_a/\kappa_v$ ($\kappa_c/\kappa_v$) significantly smaller (larger) than 1/3 and 
  the ratio $\kappa_a/\kappa_c$ much smaller than unity
  (e.g., $\kappa_a/\kappa_v=0.028$, $\kappa_c/\kappa_v=0.943$, $\kappa_a/\kappa_c=0.03$ for graphite,\cite{hanfland89}
       a most known layered material).
Thus, the PBE calculations predict BiTeI to be an extremely layered material
  whereas the PBE-D2 calculations indicate a layered, but substantially less anisotropic material.
In other words, predicting the anisotropy of BiTeI via the PBE calculations
  yields an {\it unrealistic} result,
  given that the PBE-D2 calculations are in much better agreement with the experimental compressibility data.
It appears that
  the inclusion of van der Waals attraction, which was motivated by the quasi-layered structure of BiTeI,
  leads to a much less anisotropic crystal structure.
Note that this issue does not raise in the case of PdTeI
  where van der Waals binding plays a much less prominent role 
  and therefore the PBE and PBE-D2 calculations {\it qualitatively} yield similar results, cf. Table~\ref{tabac}.

%
%
\subsection{\label{elestr} Electronic Structure}

\begin{figure*}
  \begin{center}
    \resizebox{0.33\textwidth}{!}{%
      \includegraphics{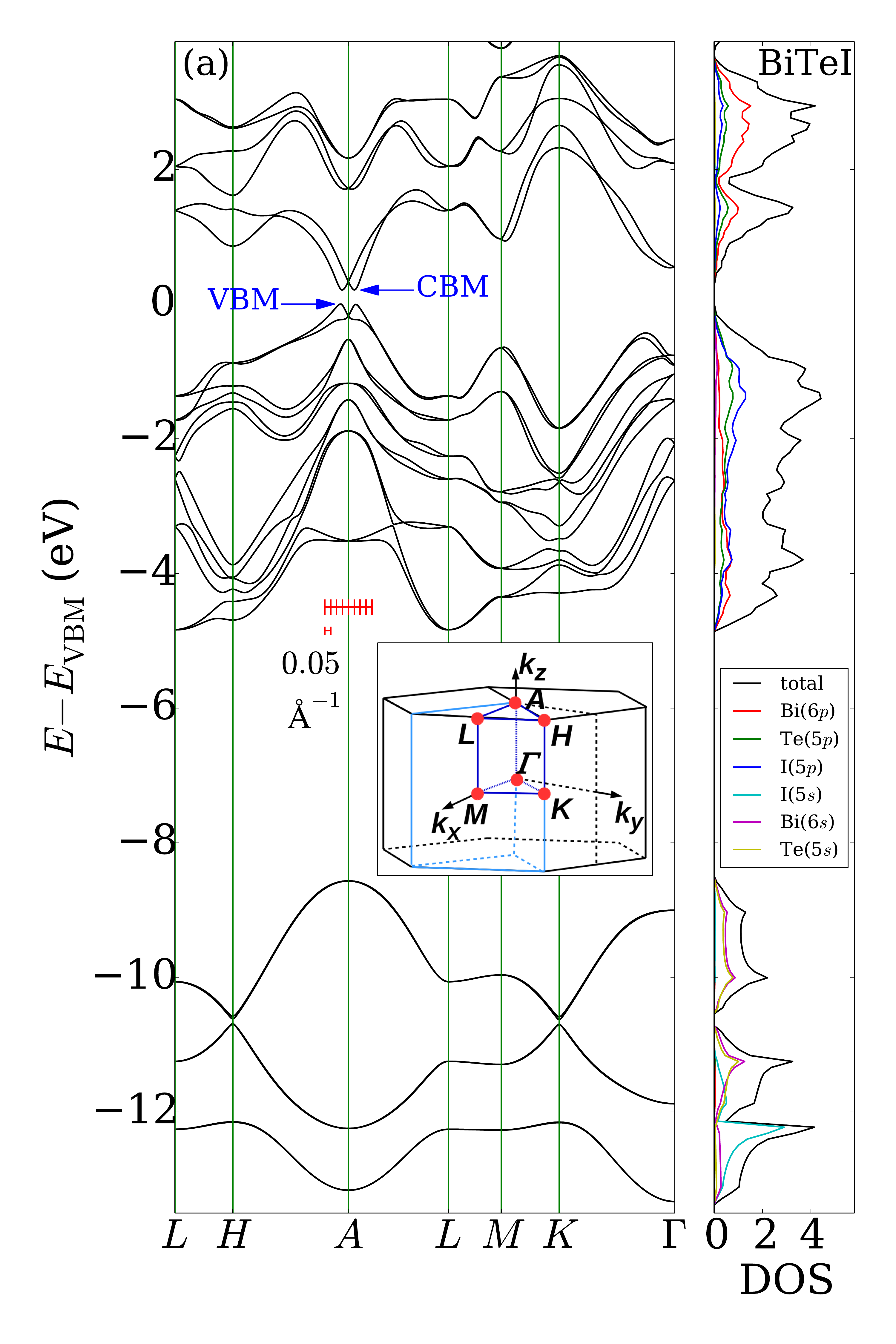}
    }
    \hspace{-0.11cm}
    \resizebox{0.33\textwidth}{!}{%
      \includegraphics{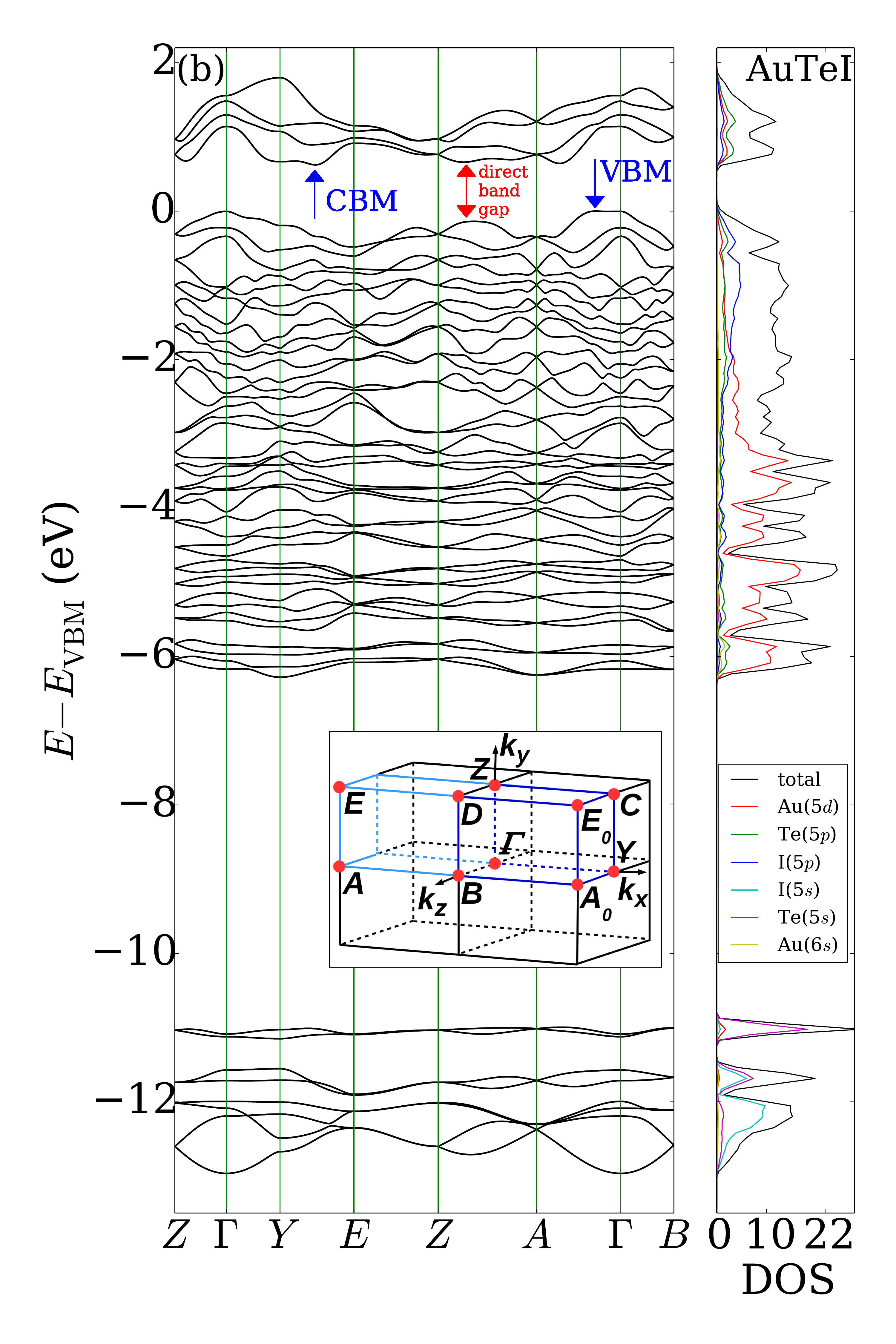}
    }
    \hspace{-0.11cm}
    \resizebox{0.33\textwidth}{!}{%
      \includegraphics{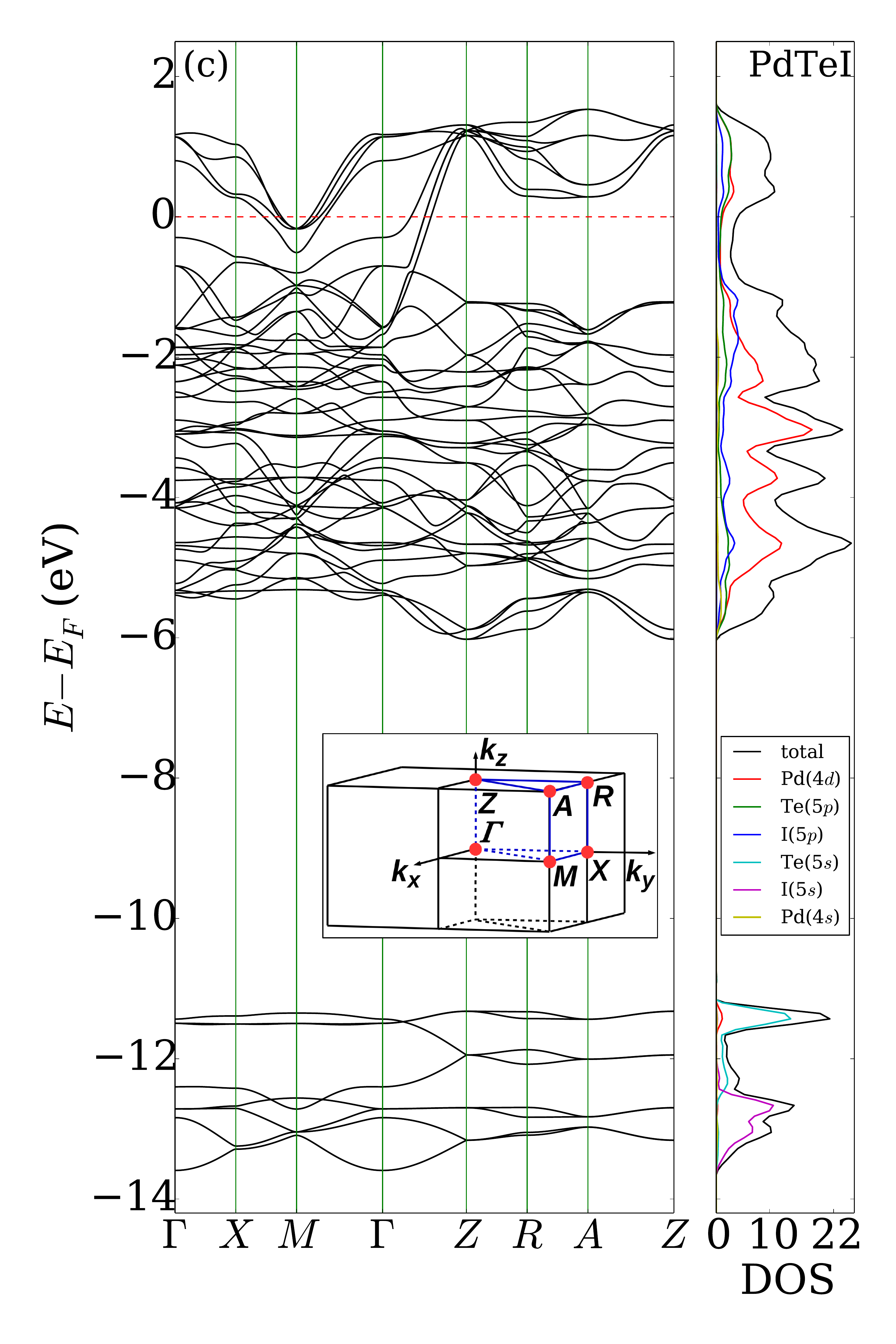}
    }
  \end{center}
  \vspace*{-0.7cm}
  \caption{(Color online) The calculated band structure (left panels in each subplots) and projected density of states (right panels in each subplots)
              of BiTeI (a), AuTeI (b), and PdTeI (c).
            The crystal structures optimized via dispersion-corrected (PBE-D2) calculations are assumed, cf. Table~\ref{tabcs}.
          }
   \label{bant}
\end{figure*}

Figure~\ref{bant}(a)-(c) displays
  the calculated band structure (left panels in each subplot) and projected density of states (right panels in each subplot)
  of the MeTeI compounds under consideration,
  obtained by using the optimized (PBE-D2) lattice parameters
  given in Table~\ref{tabcs}.
BiTeI and AuTeI appear to be narrow band gap semiconductors
  whereas PdTeI is predicted to be a metal
  with relatively low density of states at the Fermi level $E_F$.
In Fig.~\ref{bant}(a),
  Rashba-type splitting near the Brillouin zone point $A$
  is visible,
  which is in agreement with the former studies.\cite{ishizaka11,bahramy11}
This is important because
  the emergence of the bulk Rashba-type splittings
  in the calculated band structure
  {\it depends} on the reliable prediction of the BiTeI lattice parameters, cf. Fig.~4 of Ref.~\onlinecite{wang13}.
We present a comparison of the BiTeI band structures {\it with and without} SOC 
  in the Appendix~\ref{SOetkisi}, 
  which confirms that splitting near the point $A$ in Fig.~\ref{bant}(a) 
  manifests SOC-induced modification of the band edges. 
It is further shown that 
  the gradient of the crystal potential $\vec{\nabla} V(\mathbf r)$ 
  leads predominantly to an asymmetric electric field along the $c$-axis 
  that facilitates the spin-orbit coupling. 
As long as the latter gives rise to an effective magnetic field\cite{okuda13} $\vec{H}_{\rm eff}$, 
  a simple understanding is that 
  Rashba-type splitting in Fig.~\ref{bant}(a) 
  originates from the interaction of the crystal electrons with $\vec{H}_{\rm eff}$ 
  that is proportional to $\vec{\nabla} V(\mathbf r)\times {\mathbf p}$, 
  where ${\mathbf p}$ denotes the momentum operator. 
It is to be noted that a comprehensive analysis 
  has already been given by Bahramy {\it et al.} in Ref.~\onlinecite{bahramy11}, 
  revealing the origin of (giant) Rashba-type splitting in BiTeI. 

We see in Fig.~\ref{bant}(a) that 
  the lower valence band consists of the Bi, Te, and I $s$ orbitals
  whereas the upper valence band as well as lower conduction band
  are derived from  the Bi, Te, and I $p$ orbitals.
Fig.~\ref{bant}(b) shows that
  the bands of AuTeI are relatively flat (nondispersive),
  which are derived primarily from Te and I $s$ orbitals (lower valence band),
  and Te and I $s$ and Au $d$ orbitals (upper valence band as well as lower conduction band).
Au $d$ states dominate the bottom of the upper valence band.
Fig.~\ref{bant}(c) shows that
  the bands of PdTeI around the Fermi level
  are quite dispersive along not only perpendicular $\Gamma-$Z direction
  but also parallel $\Gamma$-M and X$-$M directions.
Thus our calculations indicate that
  opening a band gap owing to a possible $c$-axis-doubling distortion\cite{seo98}
  would probably not render PdTeI a semiconductor.
We see in Fig.~\ref{bant}(c) that
  the deep-lying bands consists mainly of the Te $s$ and I $s$ orbitals
  while the bands in the vicinity of the Fermi level
  are derived from the Te $p$, I $p$ and Pd $d$ orbitals.

Table~\ref{tabes} gives
  some characteristic band structure parameters for BiTeI and AuTeI semiconductors,
  for which experimental data is available:
  the band gap $E_g$  and
  the Rashba energy $E_R$ and momentum offset $k_R$ for the conduction band minimum.
As for the BiTeI band gap,
  it is interesting to point out that
  our PBE-D2 value is close the PBE value (0.242 eV) of Rusinov {\it et al.}\cite{rusinov13}
  who employed the experimental\cite{shevelkov95} lattice parameters $a$ and $c$ 
  and PBE-optimized\cite{eremeev12} internal parameters $z_{\rm Te}$ and $z_{\rm I}$ in their calculations.
Similarly,
  our PBE-D2 values for the Rashba energy and momentum offset
  are in close agreement with the respective PBE values reported in Ref.~\onlinecite{rusinov13}
  (where $E_R=$0.122 eV and $k_R=$ 0.050 \AA$^{-1}$).

It is clear from the entries of Table~\ref{tabes}
  that the calculations with the PBE functional
  yield band gaps that are {\it greater} than the respective experimental values,
  which is against the expected trend
  inasmuch as the band gap is known to be {\it underestimated} within the GGA.
Contrary to this,
  the calculated band gaps corresponding to the PBE-D2 optimized crystal structures
  are {\it smaller} than the respective experimental values,
  in line with the expected trend.
Hence, the experimental band gaps are {\it bracketed} by the respective PBE and PBE-D2 values,
  i.e.,
  \begin{eqnarray*}
    E_g^{\rm PBE-D2} <~ &E_g& ~< E_g^{\rm PBE}.
  \end{eqnarray*}
Similarly, the experimental values of $E_R$ and $k_R$
  are also {\it bracketed} by the respective PBE and PBE-D2 values (cf. Table~\ref{tabes})
  since
  \begin{eqnarray*}
    E_R^{\rm PBE}    <~ &E_R& ~< E_R^{\rm PBE-D2}, \\
    k_R^{\rm PBE-D2} <~ &k_R& ~< k_R^{\rm PBE}.
  \end{eqnarray*}
We accordingly anticipate that 
  comparative usage of the PBE and PBE-D2 calculations
  would be a useful strategy
  for predicting the band structure of layered polar semiconductors,
  which eliminates the need to employ the experimental lattice parameters.

\begin{table}
\caption{\label{tabes} 
        The calculated (PBE and PBE-D2) and experimental band structure parameters of the BiTeI and AuTeI semiconductors:
          the band gap $E_g$ (in eV) and
          the Rashba energy $E_R$ (in eV) and momentum offset $k_R$ (\AA$^{-1}$) for the conduction band minimum.
        }
\begin{ruledtabular}
\begin{tabular}{llccc}
      &                                & PBE    & PBE-D2 & Experimental \\ \hline
BiTeI &                                &        &        &              \\
      & $E_g$                 & 0.432  & 0.207  & 0.38$^a$, 
                                                  0.36$^b$,    
                                                  0.26$^c$     \\
      & $E_R$                 & 0.081  & 0.124  & 0.1$^a$      \\
      & $k_R$                 & 0.0529 & 0.0514 & 0.052$^a$    \\ 
AuTeI &                       &        &        &              \\
      & $E_g$({\rm indirect}) & 0.938  & 0.623  &              \\
      & $E_g$({\rm   direct}) & 0.938  & 0.803  & 0.9$^d$
\end{tabular}
\end{ruledtabular}
\begin{flushleft}
$^a$Ref.~\onlinecite{ishizaka11} ~~~
$^b$Ref.~\onlinecite{lee11} ~~~
$^c$Ref.~\onlinecite{sakano12} ~~~
$^d$Ref.~\onlinecite{rabenau70}
\end{flushleft}
\end{table}

\begin{figure}
  \begin{center}
    \resizebox{0.44\textwidth}{!}{%
      \includegraphics{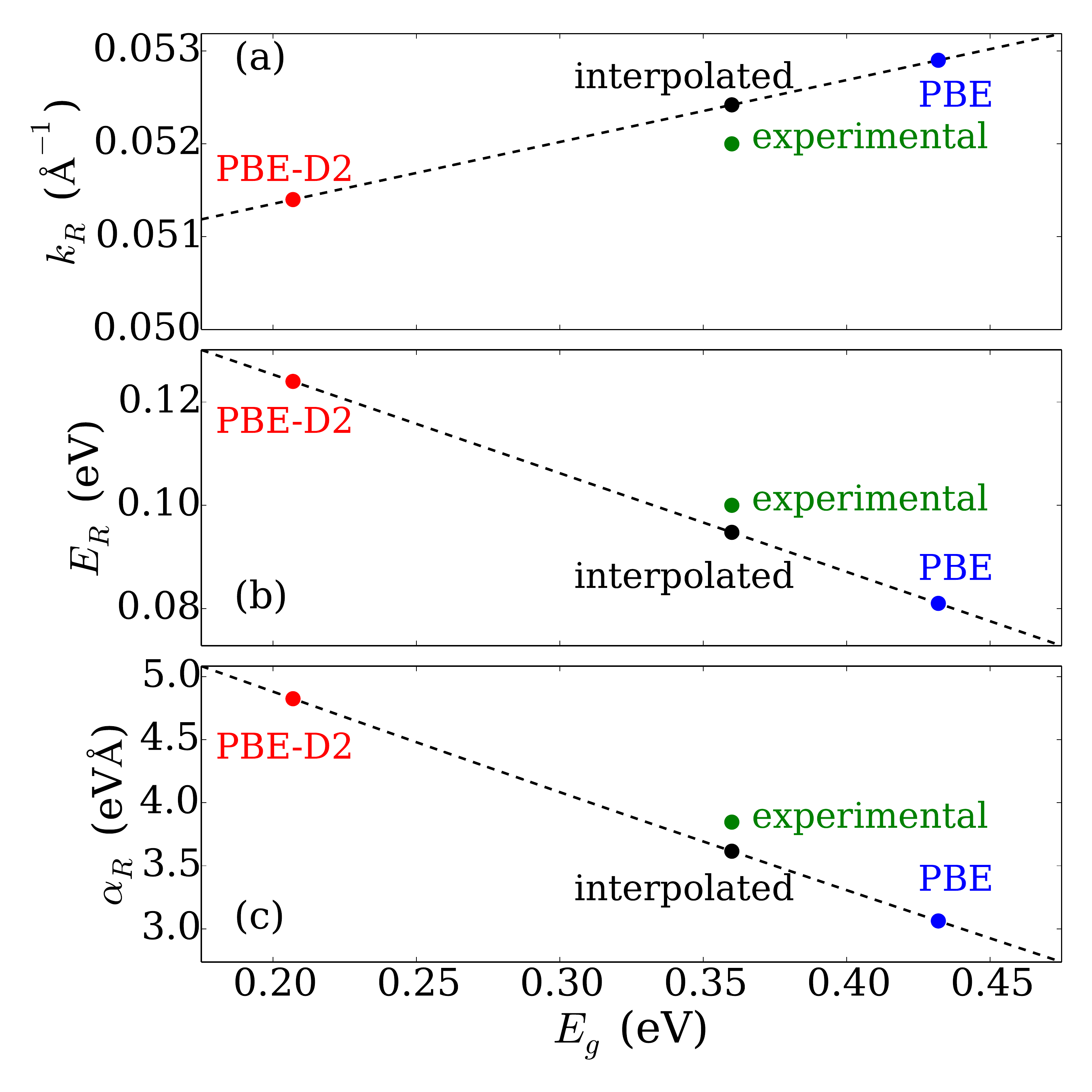}
    }
  \end{center}
  \vspace*{-0.7cm}
  \caption{(Color online)
            The momentum offset $k_R$ (a), Rashba energy $E_R$ (b), and Rashba parameter $\alpha_R$ (c)
            for the conduction band minimum of BiTeI
            versus the band gap $E_g$.
          }
   \label{RvsEg}
\end{figure}


As for the Rashba parameter $\alpha_R=2E_R/k_R$ of BiTeI,
  which represents the strength of Rashba splitting,
  the PBE-calculated and dispersion-corrected (PBE-D2) values are
   $\alpha_R=3.06$ and 4.82~eV\AA, 
  respectively,
  bracketing the experimental value\cite{ishizaka11} of $\alpha_R=3.85$~eV\AA.
It is shown in the Appendix~\ref{SOetkisi} that 
  ending up with a larger Rashba parameter within the dispersion-corrected (PBE-D2) description 
  is due to the enhancement of the asymmetric electric field $\langle -\vec{\nabla} V \rangle$ along the $c$-axis [Fig.~\ref{Efield}(b)].  

Given that the experimental values of $k_R$, $E_R$, and $\alpha_R$ are bracketed by their calculated (PBE and PBE-D2) values,
  one might attempt to interpolate their PBE-calculated and dispersion-corrected (PBE-D2) values
  in order to obtain better estimates.
The band gap $E_g$ can be used as an interpolation variable for this purpose
  since  the values of $k_R$ and $E_R$ are in almost linear correlation with the band gap values,
  as shown in Figs.~\ref{RvsEg}(a) and (b), respectively.
Note that the error due to the underestimation (PBE-D2) or overestimation (PBE) of $E_g$
  could then be corrected by using the experimental value of the band gap (i.e., $E_g=0.36$~eV)
  in the interpolation.
The dashed lines in Figs.~\ref{RvsEg}(a) and (b) passes though the calculated points,
  whose equations are given by
\begin{eqnarray}
  k_R&=&0.05002+0.00667 E_g, \nonumber \\ 
  E_R&=&0.16356-0.19111 E_g, 
  \label{kRveER}
\end{eqnarray}
respectively.
Combining the latter two equations,
  $\alpha_R$ could be obtained as a function of $E_g$,
  which is shown by the dashed line in Fig.~\ref{RvsEg}(c).
Substituting the experimental value of $E_g$ in Eq.~(\ref{kRveER}) yields
  the {\it interpolated} values of
  $k_R=$ 0.0524 \AA$^{-1}$, 
  $E_R=$0.095 eV, and
  $\alpha_R=3.62$ eV\AA~ 
  [which are marked by the black circles in Figs.~\ref{RvsEg}(a), (b) and (c), respectively]
  in 
  close  
  agreement with the respective experimental values\cite{ishizaka11}
  [marked by the green circles in Figs.~\ref{RvsEg}(a)-(c)], cf. Table~\ref{tabes}.


Finally, forasmuch as the pressure-induced closure of the BiTeI band gap 
  has been explored in a number of recent studies,\cite{bahramy12,chen13,xi13,tran14,ideue14,vangennep14} 
  we estimated the critical pressure $P_c$ at which the band gap disappears 
  (the latter being a signature of the topological phase transition). 
We obtain $P_c=3.73$~GPa 
  using our dispersion-corrected (PBE-D2) equation of state, cf. Section~\ref{subsec:eos}, 
  together with a series of band structure calculations (not shown), 
  which confirms the refined value ($P_c=3.5$~GPa) of Bahramy {\it et al.} (Refs.~\onlinecite{ideue14,bahramy12}). 
Our investigation of the variation of the BiTeI band gap and Rashba parameters with pressure 
  will be reported in detail in a separate publication. 

%
%
\section{\label{netice}Conclusion}

In summary,
  our comparative investigation of the results of semilocal (PBE) and dispersion-corrected (PBE-D2) calculations
  reveals the effect of van der Waals attractions
  on the crystal and electronic structure of metal tellurohalides BiTeI, AuTeI, and PdTeI.
We find that the prediction of crystal structure
  is systematically improved thanks to the inclusion of van der Waals dispersion.
It is shown for the compounds with a quasi-layered crystal structure, viz. BiTeI and AuTeI, that
  (i) the PBE-calculated energy versus volume curve is anomalously flat in the vicinity of equilibrium volume,
  and especially for volumes larger than the equilibrium volume, and
  (ii) the PBE-calculated equation of state has an excessively steep slope in the low-pressure region,
  which are also fixed in the dispersion-corrected (PBE-D2) calculations.
Our analysis based on the computation of
  the volume and axial compressibilities
  shows that predicting the anisotropy of BiTeI via the semilocal calculations
  yields an {\it unrealistic} result
  whereas the results of the PBE-D2 calculations agree with the experimental compressibility data.
Our calculations render that 
  BiTeI and AuTeI are narrow band gap semiconductors
  with Rashba-type spin-splitting at the band edges and
  with an indirect band gap, respectively.
PdTeI, on the other hand, is predicted to be a metal with relatively low density of states at the Fermi level.
It is notable that 
  the band gaps computed by using the optimized crystal structures from the PBE (PBE-D2) calculations
  are found to be {\it greater} ({\it smaller}) than the respective experimental values,
  which is against (in line with) the expected trend
  inasmuch as the band gap is known to be {\it underestimated} within the GGA.
We also find that the Rashba parameter, Rashba energy and momentum offset of BiTeI are {\it bracketed} by the respective values
  obtained via the semilocal (PBE) and dispersion-corrected (PBE-D2) calculations.
Specifically, a larger value for the Rashba parameter is obtained in the PBE-D2 calculations,
  which could be attributed to the reduction of the band gap caused by modification of the crystal structure
  owing to the inclusion of van der Waals dispersion terms.
Excellent agreement with the experimental data
  for the Rashba parameter, Rashba energy and momentum offset of BiTeI
  is obtained {\it via} interpolation of the calculated (PBE and PBE-D2) values.

%
%
\begin{acknowledgments}
We thank X. Xi and Y. Chen for providing the experimental data used in Figs.~\ref{uyarla}, \ref{kunc}, and S2.
The drawings in Fig.~\ref{atostr} were produced by the visualization software VESTA.\cite{momma11}
The numerical calculations reported here were carried out at the High Performance and Grid Computing Center (TRUBA Resources) of TUBITAK ULAKBIM.
\end{acknowledgments}

%
%
\appendix*

\renewcommand{\thefigure}{A\arabic{figure}}
\renewcommand{\thetable}{A}

\setcounter{figure}{0}
\setcounter{table}{0}

\section{\label{SOetkisi} Spin-Orbit-Induced Modification of the BiTeI Band Edges} 

\begin{figure}
  \begin{center}
    \resizebox{0.45\textwidth}{!}{%
      \includegraphics{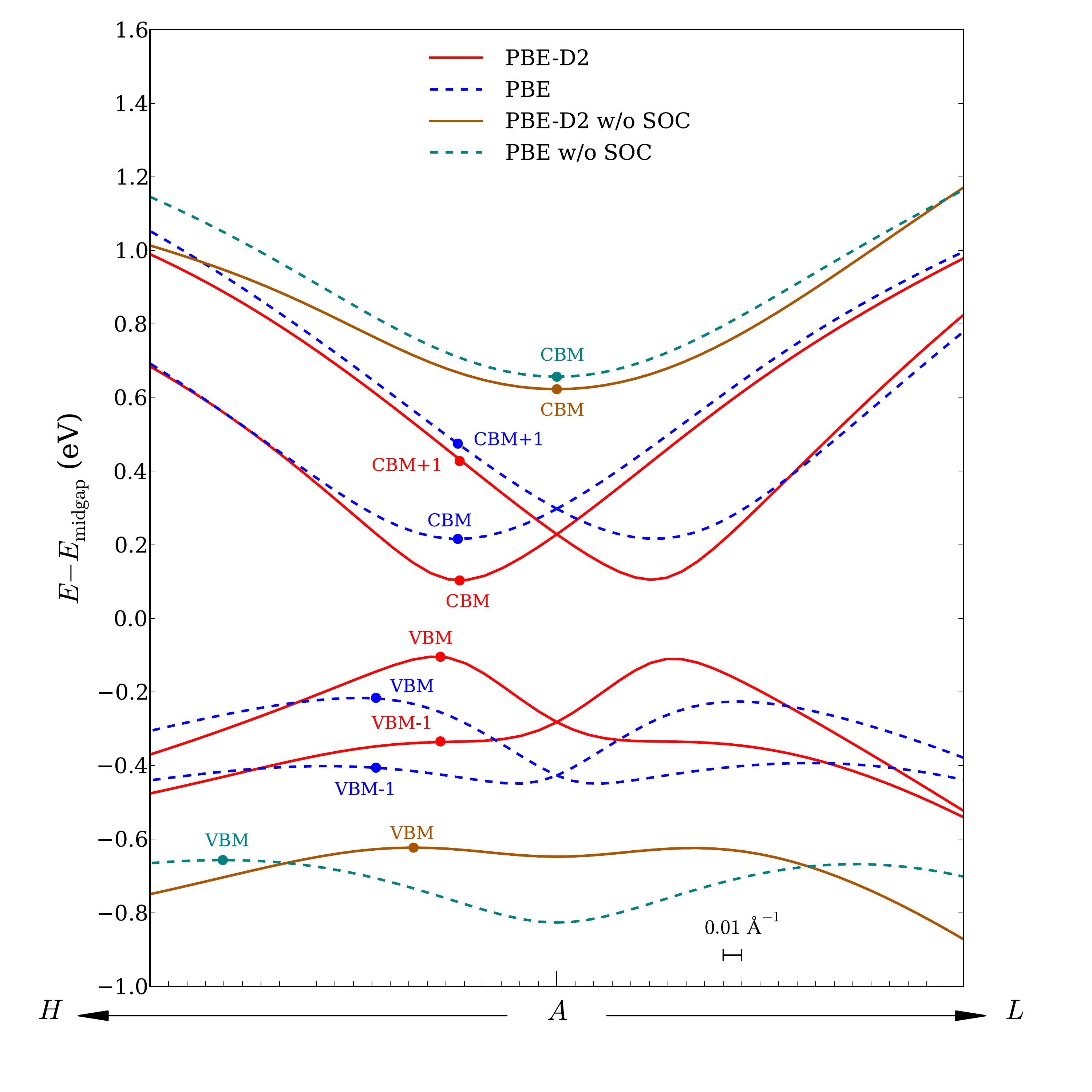}
    }
  \end{center}
  \vspace*{-0.5cm}
  \caption{(Color online)
           The band structure of BiTeI in the vicinity of the band edges,
           calculated with and without SOC and dispersion correction.
           The lower conduction (upper valence) band states labeled as CBM and CBM$+1$ (VBM and VBM$-1$) are marked by the solid circles.
          }
   \label{soimod}
\end{figure}

\begin{table*}
\caption{\label{tabprocar}
        The projected wave function character of
        the lower conduction (CBM and CBM$+1$) and upper valence (VBM and VBM$-1$) band states, cf. Fig.~\ref{soimod}.
        The entry values represent, in \%, the angular-momentum-resolved ($s$, $p_x$, $p_y$, $p_z$) contributions
        from all (Bi, Te, or I) atoms.
        The latter are rounded within 1\%, and the contributions less than 1\% are ignored.
        }
\begin{ruledtabular}
\begin{tabular}{llrrrrrrrrrrrr}
        &                & Bi  &       &       &       & Te  &       &       &       & I   &       &       &       \\
State   &                & $s$ & $p_x$ & $p_y$ & $p_z$ & $s$ & $p_x$ & $p_y$ & $p_z$ & $s$ & $p_x$ & $p_y$ & $p_z$ \\ \hline
CBM$+1$ &                &     &       &       &       &     &       &       &       &     &       &       &       \\
        & PBE-D2         &   2 &    13 &    10 &    22 &   2 &    14 &    15 &     4 &   1 &       &       &    14 \\
        & PBE            &     &    11 &     9 &    30 &   3 &    14 &    17 &     1 &   2 &     1 &       &     9 \\
CBM     &                &     &       &       &       &     &       &       &       &     &       &       &       \\
        & PBE-D2 w/o SOC &     &       &       &    69 &  12 &       &       &       &   6 &       &       &    11 \\
        & PBE-D2         &     &    12 &    16 &    28 &   5 &    14 &    10 &       &   2 &     2 &     3 &     6 \\
        & PBE            &     &    10 &    13 &    35 &   5 &    13 &     8 &       &   2 &     2 &     5 &     4 \\
        & PBE w/o SOC    &     &       &       &    72 &  10 &       &       &       &   6 &       &       &     9 \\
VBM     &                &     &       &       &       &     &       &       &       &     &       &       &       \\
        & PBE-D2 w/o SOC &  16 &       &       &     1 &     &     3 &     9 &    44 &     &       &       &    24 \\
        & PBE-D2         &  14 &       &       &     2 &     &     6 &    12 &    37 &     &     1 &     1 &    23 \\
        & PBE            &  11 &       &       &     4 &     &     9 &    18 &    34 &     &     1 &     1 &    16 \\
        & PBE w/o SOC    &  13 &       &       &     5 &     &     8 &    17 &    33 &     &     2 &     4 &    16 \\
VBM$-1$ &                &     &       &       &       &     &       &       &       &     &       &       &       \\
        & PBE-D2         &  13 &     4 &     2 &     1 &     &     7 &    10 &    35 &   1 &     6 &     3 &    14 \\
        & PBE            &  10 &     4 &     2 &       &     &    13 &    14 &    32 &     &     5 &     4 &    13 \\
\end{tabular}
\end{ruledtabular}
\end{table*}

\begin{figure}
  \begin{center}
    \resizebox{0.4\textwidth}{!}{%
      \includegraphics{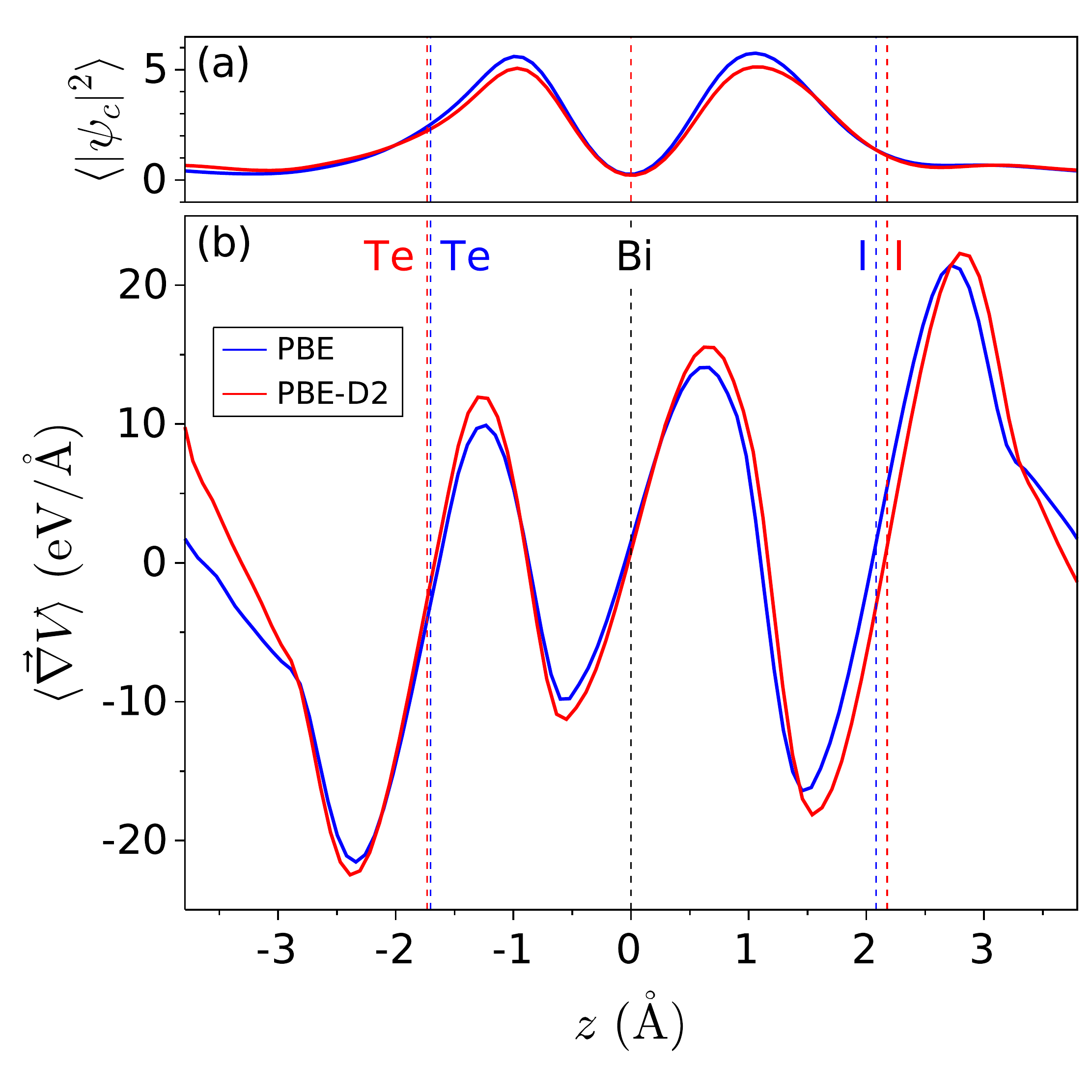}
    }
  \end{center}
  \vspace*{-0.6cm}
  \caption{(Color online)
            The $ab$ planar average of $\left | \psi_c(\mathbf r) \right |^2$ (a) and
            $\vec{\nabla} V(\mathbf r)$ (b) as a function of distance $z$ along the $c$-axis of BiTeI,
            obtained via semilocal (PBE) and dispersion-corrected (PBE-D2) calculations.
            The location of the unary Bi, Te, and I layers are marked by the vertical dashed lines.
          }
   \label{Efield}
\end{figure}

A comparison of the calculated band structures with and without SOC 
  has already been presented by Bahramy {\it et al.} in Ref.~\onlinecite{bahramy11}
  where the PBE-optimized values for $z_{\rm Te}$ and $z_{\rm I}$ were employed
  together with the {\it experimental} values for $a$ and $c$.
Here we present a similar comparison in Fig.~\ref{soimod}
  where {\it no} experimental data are used.
The lower conduction (upper valence) band states
  are marked by the solid circles
  labeled as ``CBM'' and ``CBM$+1$'' (``VBM'' and ``VBM$-1$'') in Fig.~\ref{soimod}.
Table~\ref{tabprocar} gives the projected wave function character of these states.
It is seen in Fig.~\ref{soimod} that Rashba-type splitting
  is absent in the band structures calculated {\it without} SOC (marked as ``PBE-D2 w/o SOC'' and ``PBE w/o SOC'')
  in both cases {\it with and without} the dispersion correction.
When SOC is taken into account,
  Rashba-type splitting occurs on both (conduction and valence) band edges,
  cf. the curves marked as ``PBE-D2'' and ``PBE'' in Fig.~\ref{soimod}.
It is thus clear that Rashba-type splitting in Fig.~\ref{soimod} [and Fig.~\ref{bant}(a)]
  originates from spin-orbit interaction.
Thanks to the latter,
  the band edge states have nonvanishing contributions from $p_x$ and $p_y$ orbitals
  in comparison to the cases without SOC,
  which could be seen
  via comparative inspection of rows in Table~\ref{tabprocar}.
Note, for example, that when SOC is ignored
  the character of CBM wave function $\psi_c$ is mostly of Bi-$p_z$ by 72 \% (PBE-D2) [69\% (PBE)]
  and contribution from $p_x$ and $p_y$ orbitals is negligible.
On the other hand,
  the contribution from Bi-$p_z$ is reduced to 35 \% (PBE-D2) [28\% (PBE)]
  at the expense of nonvanishing contributions from $p_x$ and $p_y$ orbitals
  when SOC is taken into account.
Thus, as demonstrated in Fig.~\ref{Efield}(a),
  the character of CBM wave function would facilitate formation of a two-dimensional electron gas
  {\it once} the Fermi level is {\it above} the conduction band minimum.
Note that the latter is usually the case with the samples used in recent experimental studies, e.g., Refs.~\onlinecite{ishizaka11,kanou13,vangennep14}.
It is seen in Fig.~\ref{Efield} that the conduction electrons
  are {\it really} subject to an asymmetric electric field $\langle -\vec{\nabla} V \rangle$ along the $c$-axis [Fig.~\ref{Efield}(b)]
  as mentioned in Section~\ref{elestr},
  which are to a significant extent confined
  in the interplanar regions (i) between Te and Bi layers as well as (ii) between the Bi and I layers [Fig.~\ref{Efield}(a)].

Figure~\ref{Efield}(b) also shows that the asymmetric electric field along the $c$-axis
  is substantially enhanced within the dispersion-corrected (PBE-D2) description (the red curve),
  compared to the PBE-calculated (blue) curve.
Note, for comparison, that there is an insignificant change
  in the $ab$ planar average of $\vec{\nabla} V$, cf. Fig.~S3 (Ref.~\onlinecite{supmat}).
Clearly, the enhancement of the asymmetric electric field along $c$-axis
  leads to stronger SOC.
Hence the value of Rashba parameter $\alpha_R$
   obtained via the dispersion-corrected (PBE-D2) calculation
   turns out to be greater than the respective PBE-calculated value.

Going back to Fig.~\ref{soimod},
  it is interesting to point out that BiTeI turns out to be an indirect semiconductor
  in all calculations presented in Fig.~\ref{soimod},
  which is in line with the former studies.\cite{kulbachinskii10,lee11,sakano12}
The difference $k_{\rm CBM}-k_{\rm VBM}$
  is largest [0.181 \AA$^{-1}$ for PBE w/o SOC]
  when SOC and dispersion correction are both ignored,
  which is reduced to 0.044 \AA$^{-1}$ (0.077 \AA$^{-1}$)
  via inclusion of SOC (dispersion correction).
Taking SOC and dispersion correction into account together yields 
  the smallest difference $k_{\rm CBM}-k_{\rm VBM}=0.010$ \AA$^{-1}$ (PBE-D2).
Similarly the band gap $E_g$ is reduced from 1.313 eV (PBE w/o SOC) to 1.246 eV (PBE-D2 w/o SOC)
  owing to the dispersion correction,
  which is further reduced to 0.207 eV (PBE-D2) via inclusion of SOC.
Thus, the reduction in the value of $E_g$ is largely due to SOC (rather than the dispersion correction).
Accordingly, strong (weak) SOC implies a smaller (larger) band gap,
  which corresponds to a greater (smaller) value of the Rashba parameter.
Hence $E_g$ and $\alpha_R$ are roughly inversely proportional to each other,
  which is also noted in a recent study\cite{zhou14} on the (BiTeI)$_m$(Bi$_2$Te$_3$)$_n$ heterostructures.

%
%

%
%
%

\clearpage

\section*{Supplemental Material}

\setcounter{page}{1}
\setcounter{figure}{0}
\setcounter{table}{0}
\renewcommand{\thepage}{S-\arabic{page}}
\renewcommand{\thefigure}{S\arabic{figure}}
\renewcommand{\thetable}{S\arabic{table}}

\begin{figure*}
  \begin{center}
    \resizebox{0.88\textwidth}{!}{%
      \includegraphics{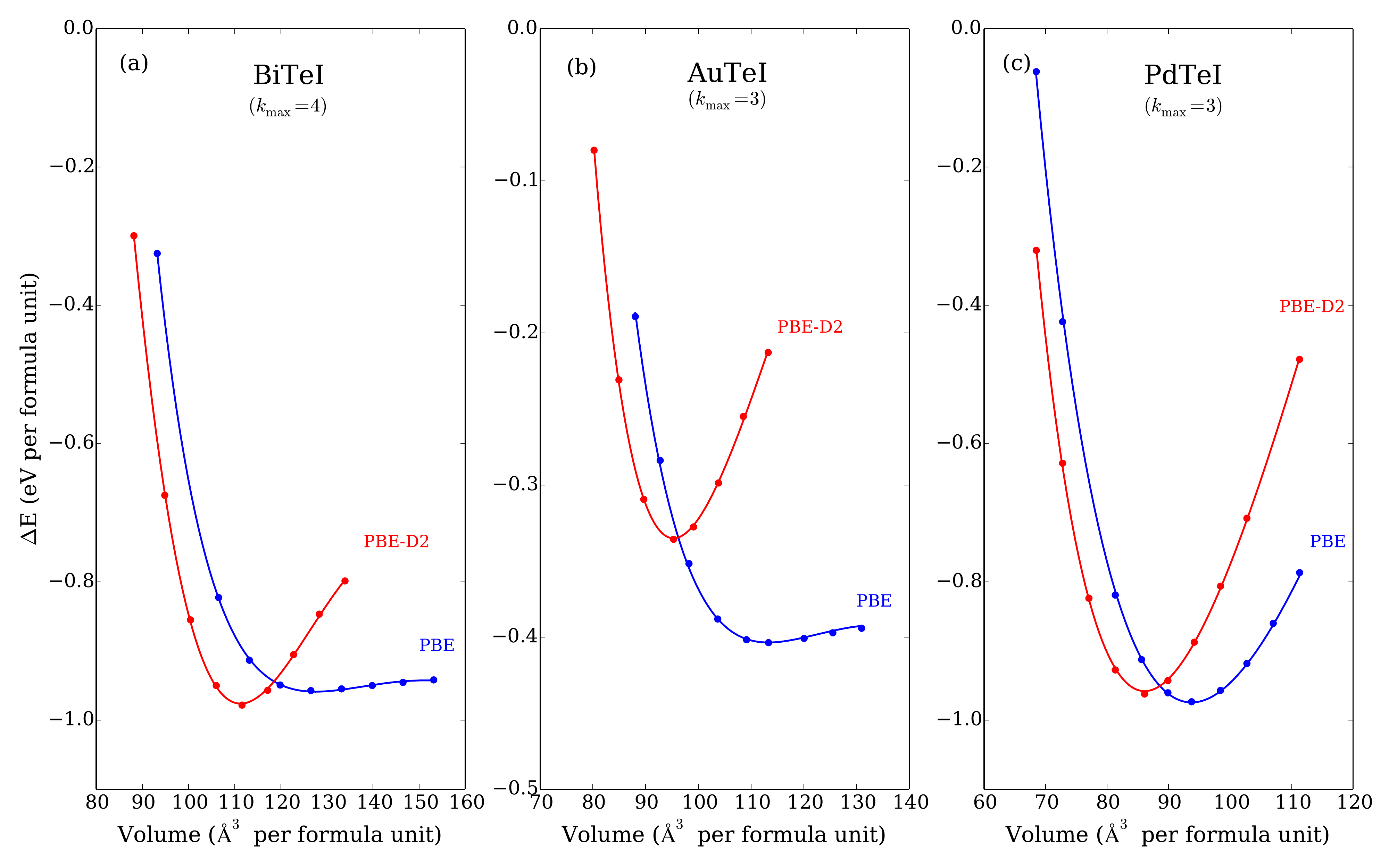}
    }
  \end{center}
  \caption{(Color online)
           The plot of the energy difference $\Delta E=E-(E_{\rm Me}+E_{\rm Te}+E_{\rm I})$, cf. Eq.~(1),
             versus the volume for BiTeI (a), AuTeI (b), and PdTeI (c).
           Note that $k_{\rm max}$ is reduced to 4 for BiTeI, and 3 for AuTeI and PdTeI.
           It is discernible that a good BM fit is achieved with these values of $k_{\rm max}$.
          }
   \label{BM433}
\end{figure*}

\begin{figure*}
  \begin{center}
    \resizebox{0.88\textwidth}{!}{%
      \includegraphics{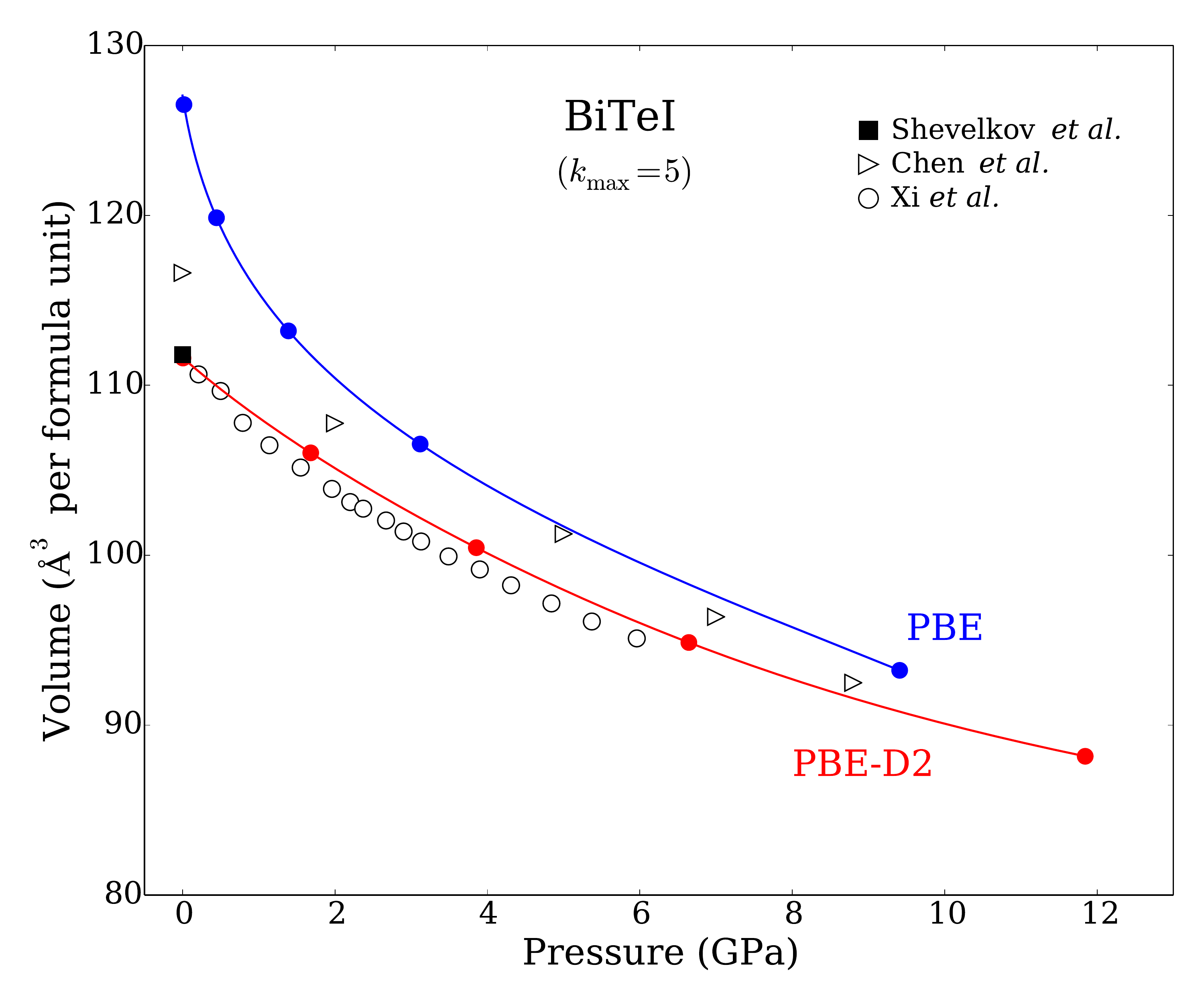}
    }
  \end{center}
  \caption{(Color online)
           The calculated and experimental pressure-volume curves for BiTeI.
           The PBE-calculated and dispersion-corrected (PBE-D2) curves are in blue and red, respectively.
           The empty circles and triangles represent the experimental data
             provided by X. Xi (Ref.~3) and Y. Chen (Ref.~14), respectively.
           The filled square marks the experimental value of the equilibrium volume
             measured by Shevelkov {\it et al.} (Ref.~15).
          }
   \label{VPexp}
\end{figure*}

\begin{figure*}
  \begin{center}
    \resizebox{0.88\textwidth}{!}{%
      \includegraphics{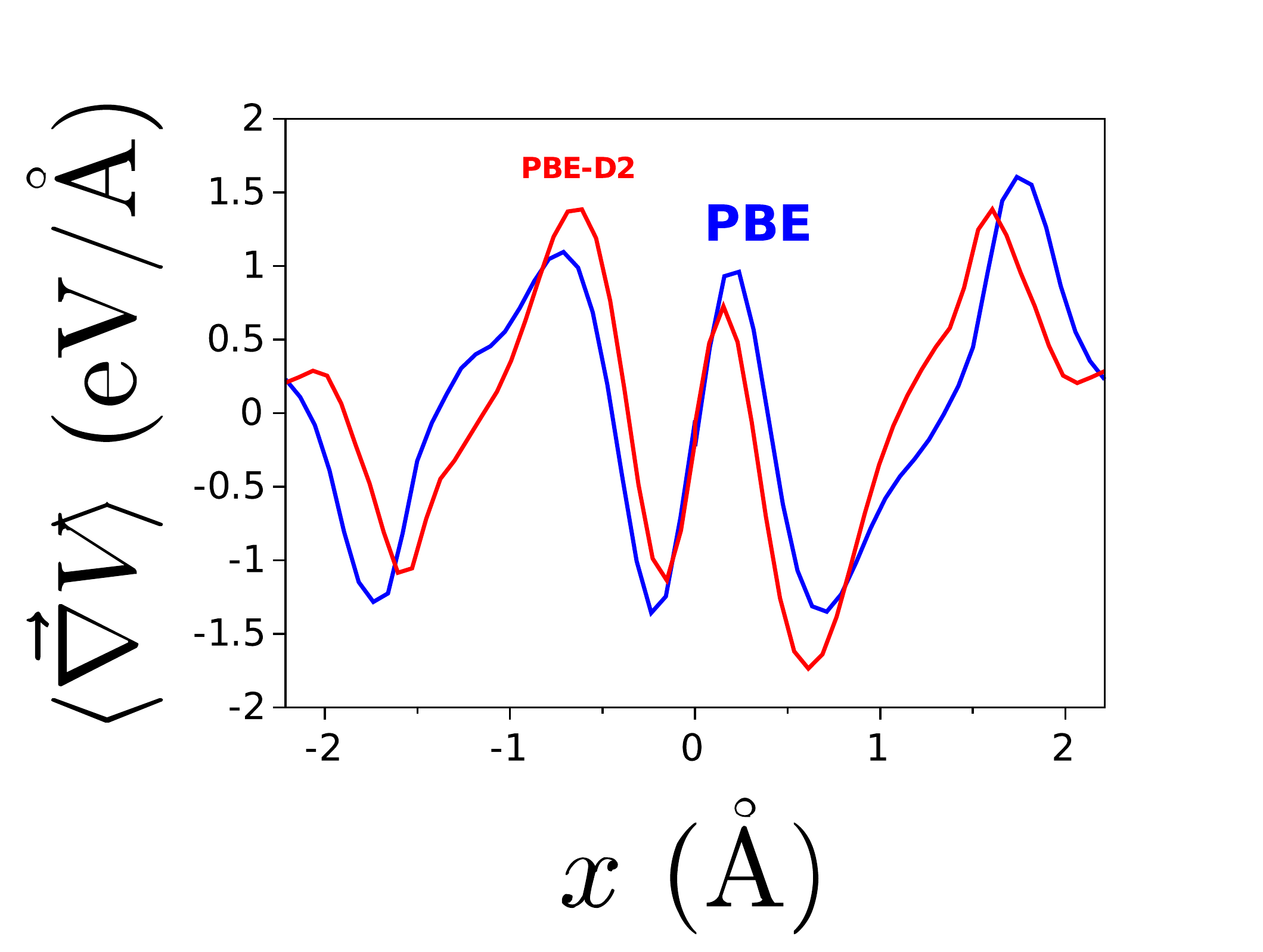}
    }
  \end{center}
  \caption{(Color online)
            The $bc$ planar average of the gradient of the BiTeI crystal potential
            as a function of distance $x$ along the $a$-axis,
            obtained via semilocal (PBE) and dispersion-corrected (PBE-D2) calculations.
          }
   \label{delVbc}
\end{figure*}

\begin{itemize}
\item Table~\ref{tabbl} lists
  the calculated and experimental bond lengths and bond angles
  in the MeTeI (Me=Bi, Au, Pd) crystals.
\item Figure~\ref{BM433} displays
  a plot of the energy difference $\Delta E$ versus the volume,
  and the BM fits with $k_{\rm max}=4$ for BiTeI and $k_{\rm max}=3$ for AuTeI and PdTeI.
\item Figure~\ref{VPexp}
  shows a comparison of the calculated and experimental pressure-volume curves for BiTeI.
\item Figure~\ref{delVbc} 
  displays the $bc$ planar average of the gradient of the BiTeI crystal potential 
  as a function of distance $x$ along the $a$-axis. 
\end{itemize}

\clearpage

%
%
\begin{table*}
\caption{\label{tabbl} The calculated (PBE and PBE-D2) and experimental bond lengths $d$ (in \AA) and bond angles $\theta$ (in $^\circ$) of the MeTeI crystals.
                       The experimental values are taken from Refs.~15,
                                                                    18, and 
                                                                    19 for BiTeI, AuTeI, and PdTeI, respectively.
                       The labeling of bond lengths and angles are therefore the same as in the latter references.
        }
\begin{ruledtabular}
\begin{tabular}{llrrr}
      &                                                  & PBE     & PBE-D2  & Experimental  \\\hline
BiTeI &                                                  &         &         &         \\
      & $d_\text{Bi-Te}$                                 &   3.08  &   3.04  &   3.27$^a$  \\
      & $d_\text{Bi-I}$                                  &   3.31  &   3.30  &   3.04$^a$  \\
      & $\theta_\text{Te-Bi-Te}$                         &  92.03  &  89.55  &  83.05$^b$  \\
      & $\theta_\text{Te-Bi-I }$                         & 174.45  & 174.10  & 174.43$^b$  \\
      & $\theta_\text{Te-Bi-I }$                         &  91.82  &  94.63  &  92.79$^b$  \\
      & $\theta_\text{I-Bi-I  }$                         &  84.06  &  80.90  &  91.11$^b$  \\
AuTeI &                                                  &         &         &         \\
      & $d_\text{Au-I}$                                  &   2.719 &   2.711 &   2.680 \\
      & $d_\text{Au-Te}$                                 &   2.671 &   2.665 &   2.642 \\
      & $d_\text{Au-Te(I)}$                              &   2.702 &   2.694 &   2.654 \\
      & $d_\text{Au-Te(II)}$                             &   2.727 &   2.694 &   2.684 \\
      & $d_{\text{Te}\cdots\text{Te(I)}}$                &   3.337 &   3.375 &   3.235 \\
      & $d_{\text{Au}\cdots\text{I(I)}}$                 &   3.711 &   3.584 &   3.500 \\
      & $d_{\text{Au}\cdots\text{I(II)}}$                &   3.753 &   3.791 &   3.633 \\
      & $d_{\text{Au}\cdots\text{Au(I)}}$                &   4.212 &   4.163 &   4.193 \\
      & $d_{\text{Au}\cdots\text{Au(II)}}$               &   4.143 &   4.034 &   4.000 \\
      & $d_{\text{Au}\cdots\text{Au(III)}}$              &   3.998 &   3.700 &   3.843 \\
      & $\theta_\text{I-Au-Te}$                          & 160.89  & 158.97  & 160.26  \\
      & $\theta_\text{I-Au-Te(I)}$                       &  85.99  &  84.27  &  86.17  \\
      & $\theta_\text{I-Au-Te(II)}$                      & 100.78  & 100.79  & 100.09  \\
      & $\theta_\text{Te-Au-Te(I)}$                      &  76.77  &  78.07  &  75.29  \\
      & $\theta_\text{Te-Au-Te(II)}$                     &  98.02  &  98.44  &  99.40  \\
      & $\theta_\text{Te(I)-Au-Te(II)}$                  & 163.69  & 170.06  & 167.71  \\
      & $\theta_\text{Au-Te-Au(I)}$                      & 103.23  & 101.93  & 104.71  \\
      & $\theta_\text{Au-Te-Au(II)}$                     & 100.25  &  97.65  &  97.34  \\
      & $\theta_\text{Au(I)-Te-Au(II)}$                  &  94.86  &  86.75  &  92.12  \\
PdTeI &                                                  &         &         &         \\
      & $d_\text{Pd-Te$_\parallel$}$                     &   2.636 &   2.629 &   2.601 \\
      & $d_\text{Pd-Te$_\perp$}$                         &   2.881 &   2.832 &   2.844 \\
      & $d_\text{Pd-I}$                                  &   2.762 &   2.742 &   2.715 \\
      & $d_\text{Pd$\cdots$Pd}$                          &   4.047 &   3.946 &   3.952 \\
      & $d_\text{Pd$\cdots$Pd}$                          &   4.048 &   3.910 &   3.873 \\
      & $d_\text{Te$_\parallel$$\cdots$Te$_\parallel$}$  &   3.377 &   3.532 &   3.385 \\
      & $d_\text{Te$_\parallel$$\cdots$Te$_\perp$}$      &   3.727 &   3.772 &   3.706 \\
      & $d_\text{I$\cdots$I}$                            &   3.758 &   3.844 &   3.809 \\
      & $\theta_\text{Te$_\parallel$-Pd-Te$_\parallel$}$ &  79.7   &  84.4   &  81.2   \\
      & $\theta_\text{I-Pd-I}$                           &  85.74  &  89.02  &  89.06  \\
      & $\theta_\text{Te$_\perp$-Pd-Te$_\perp$}$         & 166.64  & 172.64  & 168.62  \\
\end{tabular}
\end{ruledtabular}
\begin{flushleft}
$^a$Computed by using the lattice parameters given in Tables 1 and 2 of Ref.~15 since
    in Ref.~15
    the bond lengths given in Table~3 are {\it not} consistent with Tables 1 and 2,
    as noted by Bahramy {\it et al.} (Ref.~5).\\
$^b$Note that
    in Ref.~15
    the bond angles given in Table~3 are consistent with Tables 1 and 2.
\end{flushleft}

\end{table*}

%
%

%
%

%
%

\end{document}